\documentclass[12pt]{article}
\usepackage{amssymb,amsmath}
\usepackage{amsmath}
\usepackage{graphicx}
\usepackage{subfig}
\usepackage{multirow}
\usepackage{float}
\usepackage{color}
\usepackage{cite}

\newcommand{\Comment}[1]{{}}
\definecolor{MyDarkBlue}{rgb}{0.15,0.15,0.45}
\usepackage[linktocpage=true]{hyperref}
\hypersetup{
colorlinks=true,
citecolor=MyDarkBlue,
linkcolor=MyDarkBlue,
urlcolor=MyDarkBlue,
}

\def\sech{\mathop{\rm sech}\nolimits}

\begin{document}

\begin{titlepage}

\vspace{-4cm}

\title{The Worldvolume Action of Kink Solitons in AdS Spacetime {\LARGE \\[.5cm]  }}
                       
\author{{Justin Khoury, Burt A.~Ovrut, James Stokes} \\[5mm]
 {\it  Department of Physics and Astronomy} \\
 {\it University of Pennsylvania} \\
{\it Philadelphia, PA 19104---6396}\\[4mm] }

\date{}

\maketitle

\begin{abstract}
\noindent
\let\thefootnote\relax\footnotetext{jkhoury@sas.upenn.edu,~~~ovrut@elcapitan.hep.upenn.edu,\\
\indent ~~~stokesj@sas.upenn.edu}
A formalism is presented for computing the higher-order corrections to the worldvolume action of co-dimension one solitons. By modifying its potential, an explicit ``kink'' solution of a real scalar field in AdS spacetime is found. The formalism is then applied to explicitly compute the kink  worldvolume action to quadratic order in two expansion parameters--associated with the hypersurface fluctuation length and the radius of AdS spacetime respectively.
Two alternative methods are given for doing this. The results are expressed in terms of the trace of the extrinsic curvature and the intrinsic scalar curvature. In addition to conformal Galileon interactions, we find a non-Galileon term which is never sub-dominant. This method can be extended to any conformally flat bulk spacetime.

\vspace{.3in}
\noindent
\end{abstract}

\thispagestyle{empty}

\end{titlepage}

\section{Introduction}

There is a long literature on computing soliton solutions, of varying co-dimension, in both non-supersymmetric~\cite{Rajaraman:1982is,Manton:2004tk} and supersymmetric/superstring \cite{Witten:1978mh,Duff:1994an,Stelle:1996tz,Antunes:2002hn} theories of physical interest. This was followed, in each of these contexts, by computations of the lowest order Dirac-Born-Infeld (DBI) actions on the worldvolumes of these solitons~\cite{Aganagic:1996nn,Adawi:1997sq,Sorokin:1999jx,Howe:2000vk,Derendinger:2000gy,George:2009jn}. More recently, there has been interest in extending these calculations to include higher-dimensional operators, involving both extrinsic and intrinsic curvature, in these effective actions. This has been carried out with differing techniques, using both probe and back-reacted geometries, for bosonic~\cite{Gregory:1990pm,Larsen:1993yz,Barrabes:1993cn,Carter:1994ag,Arodz:1997bm,Bonjour:2000ca,Burnier:2008ke} and supersymmetric branes \cite{Bachas:1999um,Fotopoulos:2001pt,Wyllard:2001ye,Wyllard:2000qe,Andreev:1988cb,Bilal:2001hb,Howe:2001wc,Cheung:2004sa,Belyaev:2010as}.

Apart from the inherent interest in computing these higher-order corrections to the effective actions, the advent of Galileon theories has led to renewed interest in higher-derivative terms.   First discovered in the context of the decoupling limit of the Dvali-Gabadadze-Porrati (DGP) brane-world scenario~\cite{Dvali:2000hr}, where the Galileon scalar is related to the brane bending mode, Galileon theories have since been generalized~\cite{Nicolis:2008in}. Galileons have two remarkable properties: first, despite the fact that their interactions are higher-derivative, the corresponding equations of motion are nevertheless second order, and second, these interactions possess extended non-linearly realized symmetries.  
The original Galileons of \cite{Nicolis:2008in} possess another important property that justifies treating them separately from the other possible non-Galileon higher-derivative terms which possess the same symmetries.  There can be non-linear solutions and regions of momentum and field space for which Galileon terms are important relative to the kinetic terms, and yet the non-Galileon terms are unimportant, allowing us to work with only the finitely many non-linear Galileon terms rather than the whole effective field theory expansion \cite{Nicolis:2004qq,Endlich:2010zj}.  
Furthermore, Galileons can lead to stable (that is, ghost-free) violations of the Null Energy Condition~\cite{Creminelli:2006xe,Nicolis:2009qm}, akin to the ghost condensate~\cite{ArkaniHamed:2003uy}. In fact, supersymmetric condensates naturally give rise to super-Galileons~\cite{Khoury:2010gb,Khoury:2011da}. 
This violation allows for non-singular bounces in the early universe~\cite{Creminelli:2006xe,Buchbinder:2007ad, Creminelli:2007aq, Buchbinder:2007tw,Buchbinder:2007at,Lin:2010pf,LevasseurPerreault:2011mw}
and cosmologies that expand from an asymptotically-flat past~\cite{Creminelli:2010ba}.  These interesting solutions generally require the Galileon-like terms to be important relative to the kinetic terms, so from an effective field theory point of view, if other unknown higher-order terms are to be neglected, it is crucial that the Galileons have the property described in the previous paragraph.

There are extensions of the original Galileons, the DBI Galileon, that arise naturally in describing the brane-bending mode of co-dimension one and higher brane worldvolumes~\cite{deRham:2010eu,Trodden:2011xh,Hinterbichler:2010xn}.  They arise from Lovelock terms and their boundary terms in the worldvolume actions.  The original Galileons are obtained after a certain small field limit.  Non-Lovelock terms on the worldvolume lead to non-Galileon terms.  Since the DBI Galileons arise from the point of view of a brane probing a relativistic spacetime, it is natural to ask whether there can be regions of momentum and/or field space in which the DBI Galileons are important relative to the DBI kinetic term, yet still dominant over the non-Galileon terms, as is the case for the original Galileons.  If this is the case, then in these regimes, only a finite number of higher-derivative terms in the worldvolume action would have to be computed to completely determine the interesting non-linear dynamics.
The DBI Galileons admit superluminal propagation around non-trivial solutions \cite{Goon:2010xh} which, along with other arguments, suggests that any UV completion, though not necessarily inconsistent, will not be in the form of a local Lorentz invariant quantum field theory or string theory \cite{Adams:2006sv}.
Thus, if a low-energy worldvolume theory on a brane could be derived in which there were a sharp limit where only the Galileons are important, it would mean that either the theory describing the brane does not have a UV description as a local Lorentz invariant quantum field theory/string theory, or that the arguments concerning the connection between low-energy superluminality and UV physics are somehow evaded.

Motivated by these reasons, and simply by the desire to have a consistent and general method for calculating low-energy worldvolume actions for solitons in more general spacetimes, we present here a calculation of the leading and subleading corrections to the DBI worldvolume action of a scalar kink in anti-deSitter (AdS) spacetime.  This is the first in a series of papers applying this extended method to effective string solitons of physical interest, which we hope to apply to UV complete systems like the heterotic string.
In a series of papers~\cite{Gregory:1990pm,Carter:1994ag,Bonjour:2000ca}, Gregory and collaborators presented a particularly compelling approach to the problem of computing higher-order corrections to worldvolume actions, within the context of flat space co-dimension one  ``kink'' solitons. This involves a consistent series expansion in a parameter $\epsilon$, the ratio of the kink thickness to the typical worldvolume fluctuation length. Using this formalism, the explicit worldvolume action of a probe kink in a flat background bulk space was computed~\cite{Carter:1994ag}. In this paper, we modify and extend this formalism, using it to compute to second order in $\epsilon$ (and a second parameter $\delta$) the explicit worldvolume action of a kink soliton in AdS. This is carried out in two different ways, first with respect to the original AdS metric and second using a rescaled flat metric. Both lead to mutually consistent expressions for the worldvolume action including higher-order extrinsic and intrinsic curvature terms. Although three terms are indeed conformal DBI Galileons, a fourth term involving the square of the extrinsic scalar curvature explicitly is not---nor are there any momenta for which this term is sub-dominant. 

The paper is structured as follows. In Section \ref{flat}, we briefly review the formalism presented in~\cite{Carter:1994ag} for computing the worldvolume action of a kink soliton of a real scalar field in flat spacetime. As a prelude to the AdS calculation, and to set our notation, we carry out the computation to second order in the expansion parameter $\epsilon$. In Section \ref{AdS}, anti-deSitter spacetime is introduced, and the potential energy of the real scalar field is modified so that its equation of motion admits a kink soliton of the same functional form as in flat space. We then generalize the formalism of \cite{Carter:1994ag} so as to allow a calculation of the effective action on this kink worldvolume. 
The radius of AdS space introduces, in addition to $\epsilon$, a second expansion parameter $\delta$. The extrinsic and intrinsic curvatures, the generalized solution to the scalar equation of motion and the kink soliton worldvolume action are explicitly computed to second order in both parameters. Working in AdS spacetime introduces a number of technical issues, such as the appropriate ``cut-off'' of certain integrals, which are treated in detail. 
The conformally flat metric $g_{mn}$ of AdS space leads to a non-vanishing constant extrinsic scalar curvature that greatly complicates the above analysis. In the second part of Section~\ref{AdS}, we explore, in detail, the implications of working in a rescaled flat metric $\tilde{g}_{mn}$ with respect to which the lowest order extrinsic curvature vanishes. It is shown how this simplifies the computation of the worldvolume geometric quantities, while leaving the analysis of the solution of the scalar equation of motion and the kink worldvolume action essentially the same. All these quantities are explicitly calculated to second order in $\epsilon,\delta$. Using the direct relationship between the two metrics, we then compare the results of both approaches and show that they are identical, as they must be.

In Section~\ref{worldvolume}, we analyze the worldvolume action computed in the previous section. Going to a conventional gauge, each term in the action is expressed as an explicit function of a real scalar field $\pi$--the brane-bending mode. The relationship of these results to conformal Galileons~\cite{Nicolis:2008in} is discussed. In addition to the $L_2$, $L_3$ and $L_4$ Galileons (the final Galileon $L_5$ appears at one order higher than our calculation), we find that there is a non-Galileon term proportional to the square of the extrinsic curvature scalar. Importantly, it is shown that that there is no region of momentum space for which this non-Galileon term is sub-dominant. We conclude that, although important  contributions, Galileons are not the only relevant interactions on a kink/brane worldvolume. Finally, in Appendix A we prove that up to order $\epsilon^{3}$, and to all orders with no worldvolume gradient operators, the worldvolume action can always be re-expressed purely in terms of Galileons by a specific field redefinition. However, the ``non-Galileon'' physics does not disappear ---  it is now non-trivially encoded in this field transformation.

\section{Scalar Kinks in d=5 Flat Spacetime \label{flat}}

The most general action for a real scalar field $\Phi$ with minimal kinetic term coupled to gravity in $d=5$ spacetime is given by
\begin{equation}
{\cal{S}}= \int_{M_{5}} {d^{5}x \sqrt{-g}}\left(\frac{1}{2{\kappa}_{5}^{2}}(R-2\Lambda) -\frac{1}{2}g^{mn}\partial_{m} \Phi \partial_{n}\Phi-V(\Phi)\right) \ ,
\label{1}
\end{equation}
where indices $m,n=1,\dots 5$, $g_{mn}$ is the five-dimension metric with signature $(-++++)$, $\kappa_{5}$ is the dimension $-3/2$ Newton's constant, $\Lambda$ is a cosmological constant and $V(\Phi)$ is an arbitrary potential. The associated Einstein equation is
\begin{equation}
R_{mn}-\frac{1}{2}g_{mn}R+{\Lambda}g_{mn}={\kappa_{5}^{2}}T_{mn} \ ,
\label{2}
\end{equation}
where
\begin{equation}
T_{mn}=\partial_{m} \Phi \partial_{n} \Phi -g_{mn}\left(\frac{1}{2}\partial^{p} \Phi \partial_{p} \Phi+V(\Phi) \right) \ .
\label{3}
\end{equation}
Assuming that neither the temporal/spatial gradient nor the potential of  $\Phi$ depend on the $d=5$ Planck constant, in the limit that $\kappa_{5} \rightarrow 0$ the $\Phi$ dynamics decouples from gravity. Equation \eqref{2} then becomes
\begin{equation}
R_{mn}-\frac{1}{2}g_{mn}R+{\Lambda}g_{mn}=0 \ ,
\label{4}
\end{equation}
and the dynamics of the $\Phi$ field can be consistently discussed in this background spacetime --- the so-called ``probe'' limit --- using the Lagrangian 
\begin{equation}
{\cal{L}}= -\frac{1}{2}{g}^{mn} \partial_{m} \Phi \partial_{n} \Phi-V(\Phi) \ .
\label{5}
\end{equation}

In~\cite{Carter:1994ag}, Gregory and Carter used this probe limit to compute the induced worldvolume Lagrangian of the domain wall associated with the ``kink'' solution of the $\Phi$ equation of motion
in {\it flat} spacetime. In this section, we briefly review their formalism.
Begin by setting
\begin{equation}
\Lambda=0
\label{6}
\end{equation}
in \eqref{4} and taking the background spacetime to be be flat, denoting the metric by $g_{mn}=\eta_{mn}$. In Cartesian coordinates $x^{m}$, $m=0,\dots, 4$ the metric takes the diagonal form $\eta_{mn}=(-1,1,1,1,1)$.
Now specify that
\begin{equation}
V(\Phi)=\lambda(\Phi^{2}-\eta^{2})^{2} \ ,
\label{7}
\end{equation}
where $\lambda$ is a positive constant of dimension $-1$, and $\eta$ is a constant of dimension $3/2$, both independent of the $d=5$ Planck mass. The associated field equation is
\begin{equation}
\eta^{mn}{\partial}_{m}{\partial}_{n} \Phi - 4\lambda\Phi\left(\Phi^{2}-\eta^{2}\right)=0 \ .
\label{8}
\end{equation}
Denoting the the fifth coordinate $x^{5}=z$, we seek a  solution for $\Phi$ independent of the remaining coordinates. The equation of motion \eqref{8} then reduces to
\begin{equation}
\frac{d^{2} \Phi}{dz^2} - 4 \lambda \Phi\left(\Phi^{2}-\eta^{2}\right)=0 \ .
\label{9}
\end{equation}
Demanding that $\Phi$ be positive for positive values of $z$, this has the well-known ``kink'' solution 
\begin{equation}
\Phi=\eta \phi_{(0)}\,, \quad \phi_{(0)}=\tanh(\eta\sqrt{2\lambda}z) 
\label{10}
\end{equation}
of width 
\begin{equation}
l=\frac{1}{\eta\sqrt{2\lambda}} \ .
\label{11}
\end{equation}
Since this solution is independent of the remaining coordinates, it describes a static domain wall located at $z=0$. 

We would now like to generalize this to kink solutions that depend on the remaining coordinates as well as $z$. This will be achieved as follows. Let $L$ specify the typical fluctuation length of the new solution along the remaining  
coordinates and define
\begin{equation}
\epsilon=\frac{l}{L} \ .
\label{12}
\end{equation}
We will seek solutions for which $\epsilon \ll 1$ and, hence, can be obtained from \eqref{10} by a perturbation expansion. As discussed in \cite{Carter:1994ag}, this is most easily carried out in Gaussian normal coordinates, defined as follows. Let $\Phi$ be the new solution and denote the associated defect worldsheet by $\Sigma$. Let $n^{m}$ be a unit geodesic normal vector field to $\Sigma$, and generalize $z$ to be the proper length along the integral curves of $n^{m}$. The remaining four worldsheet coordinates of $\Sigma$ will be denoted by $\sigma^{\mu}$, $\mu=0,\dots,3$. Each constant $z$ surface then has a unit normal $n_{m}$, an intrinsic metric $h_{mn}$ and an extrinsic curvature $K_{mn}$ defined by
\begin{equation}
h_{mn}=\eta_{mn}-n_{m}n_{n}\,, \quad K_{mn}=h_{m}^{p} {\nabla}_{p} n_{n}\,,
\label{13}
\end{equation}
respectively. These two quantities are not independent, satisfying the constraints
\begin{eqnarray}
&&{\cal{L}}_{n}h_{mn}=2K_{mn} \ , \label{14a} \\
&&{\cal{L}}_{n}K_{mn}=K_{mp}K^{p}_{n}\,, \label{15}
\end{eqnarray}
where ${\cal{L}}_{n}$ is the Lie derivative along the $n^{m}$ vector field. With respect to Gaussian normal coordinates, the equation of motion \eqref{8} can be written as
\begin{equation}
{\cal{L}}_{n}^{2}\Phi+K{\cal{L}}_{n}\Phi+\eta^{mn}D_{m}D_{n}\Phi-4\lambda \Phi(\Phi^{2}-\eta^{2})=0 \ ,
\label{16}
\end{equation}
where
\begin{equation}
K=h^{mn}K_{mn}\,, \quad D_{m}= h_{m}^{p}\nabla_{p}   \ .
\label{17}
\end{equation}
Scaling to dimensionless variables by setting
\begin{equation}
u=\frac{z}{l}, \quad {\sigma}^{\prime}=\frac{\sigma}{L}, \quad \Phi=\eta \phi, \quad K_{mn}=\frac{1}{L}\kappa_{mn}
\label{18}
\end{equation}
equations \eqref{14a},~\eqref{15} and \eqref{16} become
\begin{eqnarray}
&&h_{mn}^{\prime}=2\epsilon \kappa_{mn} \ , \label{19} \\
&&\kappa_{mn}^{\prime}=\epsilon \kappa_{mp}\kappa^{p}_{n}\,, \label{14} \label{20} \\
&&\phi^{\prime\prime}+\epsilon \kappa \phi^{\prime}-2 \phi(\phi^{2}-1)+\epsilon^{2}D^{m}D_{m}\phi=0\,, \label{21}
\end{eqnarray}
where $^{\prime}=\partial/\partial u$.\footnote{ For notational simplicity, we henceforth drop the prime on $\sigma^{\prime}$, the dimensionality of $\sigma$ being clear from the context.} These equations can now be solved by expanding each dimensionless quantity as a power series in $\epsilon$. That is, let
\begin{eqnarray}
&& \phi=\phi_{(0)}+\epsilon \phi_{(1)}+\frac{\epsilon^{2}}{2}\phi_{(2)}+{\cal{O}}(\epsilon^{3})\,, \label{22} \\
&& h_{mn}=h_{(0)mn}+\epsilon h_{(1)mn}+\frac{\epsilon^{2}}{2}h_{(2)mn}+{\cal{O}}(\epsilon^{3})\,, \label{23} \\
&& \kappa_{mn}= \frac{1}{\epsilon}\kappa_{(0)mn}+ \kappa_{(1)mn}+\frac{\epsilon}{2} \kappa_{(2)mn}+\frac{\epsilon^{2}}{6} \kappa_{(3)mn}+ {\cal{O}}(\epsilon^{3})\,, \label{24}
\end{eqnarray}
where each coefficient is generically a function of the coordinates $(\sigma^{\mu}, u)$. Substituting these into \eqref{19},~\eqref{20} and \eqref{21}, one obtains equations for each coefficient function order by order in $\epsilon$.\\

\noindent {\it {Order $\epsilon^{0}$}}: At this order $n_{m}=n_{5}$ and, hence, $h_{(0)mn}$ is an unspecified function of $\sigma^{\mu}$ independent of $u$. It follows that equation \eqref{19} implies $\kappa_{(0)mn}$ vanishes and  \eqref{20} is trivially satisfied. That is,
\begin{equation}
h_{(0)mn}={\hat{h}}_{(0)mn}(\sigma), \quad \kappa_{(0)mn}=0 \ .
\label{25a}
\end{equation}
Here, and henceforth, any quantity that depends only on the $\sigma^{\mu}$ coordinates with be denoted with a $``$hat$"$.
The equation of motion \eqref{21} becomes
\begin{equation}
\phi_{(0)}^{\prime\prime}-2 \phi_{(0)}\left(\phi_{(0)}^{2}-1\right)=0 \ .
\label{25}
\end{equation}
This is simply \eqref{9} written in the rescaled variable $u=z/l$ and, hence, has the solution
\begin{equation}
\phi_{(0)}=\tanh(u) \ .
\label{26}
\end{equation}
\\
\noindent {\it {Order $\epsilon^{1}$}}: At this order, it follows from \eqref{20} and then \eqref{19} that
\begin{equation}
h_{(1)mn}=2u{\hat{\kappa}}_{(1)mn}\,, \quad \kappa_{(1)mn}={\hat{\kappa}}_{(1)mn}(\sigma) \,,
\label{26a}
\end{equation}
with ${\hat{\kappa}}_{(1)mn}$ unspecified. At order $\epsilon^{1}$,~\eqref{21} becomes
\begin{equation}
\phi_{(1)}^{\prime\prime}-2(3\phi_{(0)}^{2}-1)\phi_{(1)}+{\hat{\kappa}}_{(1)}(\sigma)\phi_{(0)}^{\prime}=0 \,,
\label{27}
\end{equation}
where ${\hat{\kappa}}_{(1)}={\hat{h}}_{(0)}^{mn}{\hat{\kappa}}_{(1)mn}$ is arbitrary.
Subject to the boundary conditions that $\Phi\stackrel{u \rightarrow \pm 0}{\longrightarrow}0$, $\Phi\stackrel{u \rightarrow \pm \infty}{\longrightarrow} \pm \eta$ and, hence,
\begin{eqnarray}
&& \phi_{(0)} \rightarrow 0\,, \quad \phi_{(1)} \rightarrow 0\,, ~\dots \quad {\rm as}~ u \rightarrow \pm 0\,, \label{28} \\
&& \phi_{(0)} \rightarrow \pm 1\,, \quad \phi_{(1)} \rightarrow 0\,, ~\dots \quad {\rm as}~ u \rightarrow \pm \infty\,, \label{29} 
\end{eqnarray}
there is a unique solution of \eqref{27} given by
\begin{equation}
\phi_{(1)}={\hat{\kappa}}_{(1)}(\sigma)f(u)  \,,
\label{30}
\end{equation}
with
\begin{equation}
f(u) = -\frac{1}{4} - \frac{1}{12} \cosh(2u) + \frac{1}{3} \sech^{2} u
\pm\frac{1}{12}\big(3u\sech^2u + \sinh(2u)+3\tanh u\big) \ .
\label{31}
\end{equation}
\\
\noindent We conclude that to order $\epsilon$, and restoring the dimensionful parameters, 
\begin{equation}
\Phi=\eta~ \tanh\left(\frac{z}{l}\right)+\eta l~ {\hat{K}}_{(1)}(\sigma) f\left(\frac{z}{l}\right) + {\cal{O}}(\epsilon^{2}) \ .
\label{32}
\end{equation}
As discussed in detail in \cite{Carter:1994ag}, $\Phi$ is continuous, but not continuously differentiable, across the $z=0$ wall surface.
\\

\noindent {\it {Order $\epsilon^{2}$}}: To this order, one need only know the metric and extrinsic curvature. Solving \eqref{20} and then \eqref{19}, we find that
\begin{equation}
h_{(2)mn}=2 u^{2}{\hat{\kappa}}_{(1)mp}{\hat{\kappa}}_{(1)n}^{p}\, , \quad \kappa_{(2)mn}=u{\hat{\kappa}}_{(1)mp}{\hat{\kappa}}_{(1)n}^{p} \ .
\label{32a}
\end{equation}
\\
Note that none of the purely $\sigma$ dependent quantities --- that is, none of the hatted functions --- have been determined by the above procedure. This will remain true to any order in the 
$\epsilon$-expansion. There is a fundamental reason for this; namely,
prior to computing the worldvolume action of the domain wall, one must leave unrestrained any degrees of freedom intrinsic to the wall itself. These can only be determined by varying the worldvolume action to get the equations of motion of the wall location. It follows that $h_{mn}$ and $\kappa_{mn}$ are off-shell and, hence, arbitrary functions of the intrinsic coordinates $\sigma^{\mu}$ at this stage of the calculation.

The worldvolume effective action can now be calculated to any required accuracy in the $\epsilon$ expansion. It is given by
\begin{equation} 
S_{4}=\int_{M_{4}}{{\rm d}^{4}\sigma \sqrt{-h}|_{u=0}} \hat{\cal{L}}
\label{33}
\end{equation}
where
\begin{equation}
\hat{\cal{L}}=\int{{\rm d}z} J {\cal{L}}\,, \quad   J= \frac{\sqrt{-\eta}}{\sqrt{-h}|_{u=0}}\,,
\label{34}
\end{equation}
and ${\cal{L}}$ is the original Lagrangian density given in \eqref{5},~\eqref{7} evaluated for the solution of the equation of motion given to order $\epsilon$ in \eqref{32}. Taylor expanding $\sqrt{-\eta}$ around $u=0$ and using \eqref{19},~\eqref{20} to second order, one finds
\begin{equation}
J=1+\epsilon J_{(1)} +\frac{\epsilon^{2}}{2}J_{(2)}+{\cal{O}}(\epsilon^{3}) \ ,
\label{35}
\end{equation}
with
\begin{equation}
J_{(1)}= u {\hat{\kappa}}_{(1)}\,, \quad J_{(2)}= u^{2}\left({\hat{\kappa}}_{(1)}^{2}-{\hat{\kappa}}_{(1)mn} {\hat{\kappa}}_{(1)}^{mn}\right) \ .
\label{36}
\end{equation}
Similarly, inserting solution $\Phi$ in \eqref{32} into \eqref{5},\eqref{7}, going to dimensionless variables and using the equations of motion \eqref{25} and \eqref{27}, it follows that
\begin{equation}
{\cal{L}}={\cal{L}}_{(0)}+\epsilon {\cal{L}}_{(1)}+\frac{\epsilon^{2}}{2}{\cal{L}}_{(2)}+{\cal{O}}(\epsilon^{3}) \ ,
\label{37}
\end{equation}
where
\begin{eqnarray}
&&\quad {\cal{L}}_{(0)}=-2\lambda \eta^{4} \phi_{(0)}^{\prime 2}, \quad {\cal{L}}_{(1)}=-2 \lambda \eta^{4}(\phi_{(0)}^{\prime}\phi_{(1)})^{\prime} \label{38} \\
&& {\cal{L}}_{(2)}=-2 \lambda \eta^{4}\left( (\phi_{(0)}^{\prime}\phi_{(2)})^{\prime}+ (\phi_{(1)}^{\prime}\phi_{(1)})^{\prime} +{\hat{\kappa}}_{(1)}\phi_{(0)}^{\prime}\phi_{(1)} \right)\  . \nonumber
\end{eqnarray}
Multiplying \eqref{35} and \eqref{37} then gives
\begin{eqnarray}
&&\qquad \frac{J {\cal{L}}}{\lambda \eta^{4}}= -2\phi_{(0)}^{\prime ~2}\left( 1+\epsilon J_{(1)} +\frac{\epsilon^{2}}{2}J_{(2)} \right)+\epsilon^{2} {\hat{\kappa}}_{(1)} \phi_{(0)}^{\prime}\phi_{(1)}
\label{39} \\
&&-\epsilon \left( \phi_{(0)}^{\prime}(2 \phi_{(1)}+\epsilon \phi_{(2)})+\epsilon(\phi_{(1)}^{\prime}+2{\hat{\kappa}}_{(1)}\phi_{(0)}^{\prime}u) \phi_{(1)} \right)^{\prime}+{\cal{O}}(\epsilon^{3}) \nonumber
\end{eqnarray}
in each of the separate domains $-\infty <u<0$ and $0<u<\infty$. Note that the vanishing of $\kappa_{(0)mn}$ allows one to equate
\begin{equation}
{\hat{\kappa}}={\hat{\kappa}}_{(1)}
\label{39A}
\end{equation}
to this order in the $\epsilon$-expansion, which we do henceforth.
The above stated boundary conditions imply that when integrated over $-\infty<u<\infty$ the contribution of the total divergence term vanishes. Furthermore, $J_{(1)}$ in \eqref{36} is odd in $u$ and also does not contribute. Hence, inserting \eqref{39} into \eqref{34} using \eqref{36}, $z=lu$ and \eqref{11} one finds
\begin{equation}
{\hat{\cal{L}}}={\hat{\cal{L}}}_{(0)}+\frac{\epsilon^{2}}{2}{\hat{\cal{L}}}_{(2)}+{\cal{O}}(\epsilon^{3}) \ ,
\label{40}
\end{equation}
where
\begin{eqnarray}
&&{\hat{\cal{L}}}_{(0)}=-\frac{\eta^{2}}{l} I_{I}  \label{41} \\
&&{\hat{\cal{L}}}_{(2)}=-\frac{\eta^{2}}{l} \left({\hat{\kappa}}^{2}-{\hat{\kappa}}_{mn}{\hat{\kappa}}^{mn} \right)
I_{II}+\frac{\eta^{2}}{l} {\hat{\kappa}}^{2} I_{III} \nonumber
\end{eqnarray}
and
\begin{eqnarray}
&&I_{I}=\int_{-\infty}^{+\infty}{du~ \phi_{(0)}^{\prime~2}}=\frac{4}{3}, \quad I_{II}=\int_{-\infty}^{+\infty}{du~ u^{2} \phi_{(0)}^{\prime~2}}=\frac{\pi^{2}-6}{9} \nonumber \\
&&\qquad \qquad \qquad I_{III}=\int_{-\infty}^{+\infty}{du~ f~ \phi_{(0)}^{\prime}}=\frac{5}{18} \ . \label{42} 
\end{eqnarray}
Rewritten in dimensionful variables, truncating the expansion at order $\epsilon^{2}$ and using
the Gauss-Codazzi relation 
\begin{equation}
{\hat{R}}^{(4)}={\hat{K}}^{2}-{\hat{K}}_{m}^{n}{\hat{K}}_{n}^{m} \ ,
\label{44}
\end{equation}
the worldvolume Lagrangian is given by
\begin{equation}
{\hat{\cal{L}}}=-\frac{4\eta^{2}}{3l}\left(1+C_{I}{\hat{R}}^{(4)}+C_{II}{\hat{K}}^{2}  \right) 
\label{43}
\end{equation}
where 
\begin{equation}
C_{I}=\frac{I_{II}} {I_{I}}  \frac{l^{2}}{2}=\left(\frac{\pi^{2}-6}{24}\right) l^{2}, \quad C_{II}=-\frac{I_{III}} {I_{I}}  \frac{l^{2}}{2}=-\frac{5}{48} l^{2} \ .
\label{45}
\end{equation}
This is the result presented by Gregory and Carter \cite{Carter:1994ag}, after correcting some errors in their manuscript.

\section{Scalar Kinks in d=5 AdS Spacetime \label{AdS}}

In this section we will use, and extend, the Gregory/Carter formalism to compute the worldvolume Lagrangian of a kink domain wall in $d=5$ anti-deSitter spacetime. In this case, we choose
\begin{equation}
\Lambda < 0 \ .
\label{46}
\end{equation}
Equation \eqref{4} then admits an AdS background solution with metric  
\begin{equation}
ds^{2}=e^{\frac{2z}{{\cal{R}}}} dx^{\mu} dx^{\nu} \eta_{\mu\nu}+dz^{2} 
\label{47}
\end{equation}
where $\Lambda=-\frac{6}{{\cal{R}}^{2}}$. The associated curvature tensors are given by
\begin{equation}
R^{(5)}=-\frac{20}{{\cal{R}}^{2}}, \quad R^{(5)}_{mn}=-\frac{4}{{\cal{R}}^{2}}g_{mn}, \quad R^{(5)}_{mnpq}=-\frac{1}{{\cal{R}}^{2}} \left(g_{mp}g_{nq}-g_{mq}g_{np}  \right) \ .
\label{48}
\end{equation}
In the ``probe'' limit, the dynamics of the $\Phi$ field can be consistently discussed in this AdS background using the Lagrangian 
\begin{equation}
{\cal{L}}= -\frac{1}{2}{g}^{mn} \partial_{m} \Phi \partial_{n} \Phi-V(\Phi) \ .
\label{49}
\end{equation}
In order for the equation of motion to admit a kink solution, we must modify potential \eqref{7} to 
\begin{equation}
V(\Phi)=\lambda(\Phi^{2}-\eta^{2})^{2}+\frac{4\sqrt{2\lambda}}{{\cal{R}}}\left( \eta^{2}\Phi- \frac{1}{3} \Phi^{3} \right) \ .
\label{50}
\end{equation}
The associated field equation now becomes
\begin{equation}
g^{mn}{\nabla}_{m}{\partial}_{n} \Phi - 4\lambda\Phi(\Phi^{2}-\eta^{2})-\frac{4\sqrt{2\lambda}}{{\cal{R}}}\left( \eta^{2}-\Phi^{2} \right)   =0 \ .
\label{51}
\end{equation}
We seek a solution for $\Phi$ that depends on the fifth coordinate $z$ but is independent of the remaining coordinates. The equation of motion \eqref{51} then reduces to
\begin{equation}
\frac{d^{2} \Phi}{dz^{2}}+\frac{4}{{\cal{R}}}\frac{d \Phi}{dz} - 4 \lambda \Phi(\Phi^{2}-\eta^{2})-\frac{4\sqrt{2\lambda}}{{\cal{R}}}\left( \eta^{2}-\Phi^{2} \right) =0 \ .
\label{52}
\end{equation}
Despite the fact that we are now in AdS spacetime, \eqref{52} continues to admit the kink solution
\begin{equation}
\Phi=\eta \phi_{(0)}, \quad \phi_{(0)}=\tanh(\eta\sqrt{2\lambda}z) 
\label{53}
\end{equation}
of width 
\begin{equation}
l=\frac{1}{\eta\sqrt{2\lambda}} \ .
\label{54}
\end{equation}
Since this solution is independent of the remaining coordinates, it describes a static domain wall located at $z=0$. 

We would now like to generalize this to kink solutions that depend on the remaining coordinates as well as $z$. Specifying the typical fluctuation length along the remaining coordinates as $L$, and defining
\begin{equation}
\epsilon=\frac{l}{L} \ ,
\label{55}
\end{equation}
this will again be achieved as a perturbative expansion around \eqref{53} in the small parameter $\epsilon \ll 1$. As in the flat spacetime case, this is most easily carried out in the Gaussian normal coordinates defined in Section \ref{flat}. Each constant $z$ surface has a unit normal $n_{m}$, with an intrinsic metric $h_{mn}=g_{mn}-n_{m}n_{n}$ and extrinsic curvature $K_{mn}$ defined in \eqref{13}.
These two quantities are not independent. As in flat spacetime, the metric and extrinsic curvature continue to be related as
\begin{equation}
{\cal{L}}_{n}h_{mn}=2K_{mn} \ ,
\label{56} 
\end{equation}
where ${\cal{L}}_{n}$ is the Lie derivative along the $n^{m}$ vector field. However, in a general curved five-dimensional spacetime we note from a Gauss-Codazzi relation that
\begin{equation}
{\cal{L}}_{n}K_{mn}=K_{mp}K^{p}_{n}-R^{(5)}_{rspq}n^{s}n^{q}h^{r}_{m}h^{p}_{n} \ .
\label{57}
\end{equation}
Using the expression for $R^{(5)}_{rspq}$ in AdS spacetime given in \eqref{48} and the definition of $h_{mn}$ in \eqref{13}, we find 
\begin{equation}
R^{(5)}_{rspq}n^{s}n^{q}h^{r}_{m}h^{p}_{n}=-\frac{1}{{\cal{R}}^{2}}h_{mn}
\label{58}
\end{equation} 
and, hence, that
\begin{equation}
{\cal{L}}_{n}K_{mn}=K_{mp}K^{p}_{n}+\frac{1}{{\cal{R}}^{2}}h_{mn}  \ .
\label{59}
\end{equation}
Note that in the limit ${\cal{R}}\rightarrow \infty$, this equation reverts to the flat spacetime expression given in \eqref{15}.
In Gaussian normal coordinates, the equation of motion \eqref{51} becomes
\begin{equation}
{\cal{L}}_{n}^{2}\Phi+K{\cal{L}}_{n}\Phi+D^{m}D_{m}\Phi-4\lambda \Phi(\Phi^{2}-\eta^{2})-\frac{4\sqrt{2\lambda}}{{\cal{R}}}\left( \eta^{2}-\Phi^{2} \right) =0 \ ,
\label{60}
\end{equation}
where $K$ and $D_{m}$ are the extrinsic scalar curvature and worldvolume covariant derivative defined in \eqref{17}.
Scaling to dimensionless variables by setting
\begin{equation}
u=\frac{z}{l}, \quad \Phi=\eta \phi, \quad K_{mn}=\frac{1}{L}\kappa_{mn}
\label{61}
\end{equation}
equations \eqref{56},\eqref{59} and \eqref{60} become
\begin{eqnarray}
&&h_{mn}^{\prime}=2\epsilon \kappa_{mn} \ , \label{62} \\
&&\epsilon \kappa_{mn}^{\prime}=\epsilon^{2} \kappa_{mp}\kappa^{p}_{n} +\delta^{2} h_{mn} \ ,  \label{63} \\
&&\phi^{\prime\prime}+\epsilon \kappa \phi^{\prime}-2 (\phi-2\delta)(\phi^{2}-1)+\epsilon^{2}D^{m}D_{m}\phi=0 \label{64}
\end{eqnarray}
where $^{\prime}=\frac{\partial}{\partial u}$ and we have defined 
\begin{equation}
\delta=\frac{l}{{\cal{R}}} \ .
\label{64a}
\end{equation}
These equations can now be solved by expanding each dimensionless quantity as a power series in $\epsilon$. That is, let
\begin{eqnarray}
&& \phi=\phi_{(0)}+\epsilon \phi_{(1)}+\frac{\epsilon^{2}}{2}\phi_{(2)}+{\cal{O}}(\epsilon^{3}), \label{65} \\
&& h_{mn}=h_{(0)mn}+\epsilon h_{(1)mn}+\frac{\epsilon^{2}}{2}h_{(2)mn}+{\cal{O}}(\epsilon^{3}), \label{66} \\
&& \kappa_{mn}= \frac{1}{\epsilon}\kappa_{(0)mn}+ \kappa_{(1)mn}+\frac{\epsilon}{2} \kappa_{(2)mn}+\frac{\epsilon^{2}}{6} \kappa_{(3)mn}+ {\cal{O}}(\epsilon^{3}) \label{67}
\end{eqnarray}
where each coefficient is generically a function of the coordinates $(\sigma^{\mu}, u)$. Substituting these into \eqref{62},\eqref{63} and \eqref{64}, one obtains equations for each coefficient function order by order in $\epsilon$.

The non-zero curvature in the AdS analysis will require us to carefully examine the $\epsilon$ expansion of the $h_{mn}$ and $\kappa_{mn}$ equations. First consider the $h_{mn}$ equation \eqref{62}. Substituting \eqref{66} and \eqref{67} into \eqref{62}, we find to  order $\epsilon^{0}$ 
and $\epsilon^{1}$ that
\begin{eqnarray}
&&h^{\prime}_{(0)mn}=2\kappa_{(0)mn}\ , \label{68}  \\
&&h^{\prime}_{(1)mn}=2\kappa_{(1)mn} \label{69} 
\end{eqnarray}
respectively. Now examine the the $\kappa_{mn}$ equation \eqref{63}. Note that  this can be written as 
\begin{equation}
\epsilon \kappa_{mn}^{\prime}=\epsilon^{2} \kappa_{mp}\kappa_{qn}h^{pq}  +\delta^{2} h_{mn}
\label{lamp1} 
\end{equation}
where
\begin{equation}
h^{mq}h_{qn}=\delta^{m}_{n} \ .
\label{69a}
\end{equation}
Expanding 
\begin{equation}
h^{mn}=h_{(0)}^{mn}+\epsilon h_{(1)}^{mn}+\frac{\epsilon^{2}}{2}h_{(2)}^{mn}+{\cal{O}}(\epsilon^{3}) \ ,
\label{69b}
\end{equation}
it follows from \eqref{66}, \eqref{69a} that
\begin{equation}
h_{(0)}^{mq}h_{(0)qn}=\delta^{m}_{n} \ , \quad h_{(0)}^{mq}h_{(1)qn}+h_{(1)}^{mq}h_{(0)qn}=0 \ .
\label{69c}
\end{equation}
Substituting \eqref{66}, \eqref{67} and \eqref{69b} into the $\kappa_{mn}$ equation \eqref{lamp1}, we find to  order $\epsilon^{0}$ and  $\epsilon^{1}$ that
\begin{eqnarray}
&& \kappa^{\prime}_{(0)mn}= \kappa_{(0)mp}\kappa_{(0)qn}h_{(0)}^{pq}+\delta^{2}h_{(0)mn}\ , \label{70} \\
&&\kappa^{\prime}_{(1)mn}= \kappa_{(0)mp}\kappa_{(1)qn}h_{(0)}^{pq}+ \kappa_{(1)mp}\kappa_{(0)qn}h_{(0)}^{pq} \label{71} \\
&&\qquad  \qquad+\kappa_{(0)mp} \kappa_{(0)qn} h_{(1)}^{pq} + \delta^{2}h_{(1)mn} \ . \nonumber
\end{eqnarray}
Before continuing to the equation of motion, let us solve \eqref{68},\eqref{69} and \eqref{70},
\eqref{71}. \\

\noindent {\it Order $\epsilon^{0}$:} Since at this order $n_{m}=n_{5}$, it follows from \eqref{47} that 
\begin{equation}
h_{(0)mn}=e^{2\delta u}{\hat{h}}_{(0)mn}(\sigma) 
\label{74}
\end{equation}
with ${\hat{h}}_{(0)mn}$ unspecified.
Hence, the $h_{(0)mn}$ equation \eqref{68} gives
\begin{equation}
\kappa_{(0)mn}=\delta h_{(0)mn} \ .
\label{75}
\end{equation}
For notational consistency, we write this as
\begin{equation}
\kappa_{(0)mn}=e^{2\delta u}{\hat{\kappa}}_{(0)mn}(\sigma), \quad {\hat{\kappa}}_{(0)mn}(\sigma)=\delta {\hat{h}}_{(0)mn}(\sigma) \ .
\label{lamp5}
\end{equation}
Using the first relation in \eqref{69c}, we find that the $\kappa_{(0)mn}$ equation \eqref{70} is trivially satisfied.\\

\noindent {\it Order $\epsilon^{1}$:} Substituting \eqref{75} into the order $\epsilon^{1}$ $\kappa_{(1)mn}$ equation \eqref{71}, and recognizing that the second expression in \eqref{69c} implies
\begin{equation}
h_{(0)mp}h_{(0)qn}h_{(1)}^{pq}=-h_{(1)mn} \ ,
\label{76}
\end{equation}
we find that
\begin{equation}
\kappa^{\prime}_{(1)mn}=2\delta \kappa_{(1)mn} \ .
\label{77}
\end{equation}
This is solved by
\begin{equation}
\kappa_{(1)mn}=e^{2\delta u}{\hat{\kappa}}_{(1)mn}(\sigma)
\label{78}
\end{equation}
with unspecified ${\hat{\kappa}}_{(1)mn}(\sigma)$. Putting this result into the order 
$\epsilon^{1}$ $h_{(1)mn}$ equation \eqref{69} gives
\begin{equation}
h^{\prime}_{(1)mn}=2e^{2\delta u}{\hat{\kappa}}_{(1)mn}(\sigma) \ .
\label{79}
\end{equation}
This can be integrated to 
\begin{equation}
h_{(1)mn}=\frac{1}{\delta}\left(\kappa_{(1)mn}-{\hat{\kappa}}_{(1)mn}(\sigma)  \right) \ .
\label{80}
\end{equation}
To summarize: we have found that
\begin{eqnarray}
&& h_{(0)mn}=\frac{1}{\delta}\kappa_{(0)mn}, \qquad h_{(1)mn}=\frac{1}{\delta}\left(\kappa_{(1)mn}-{\hat{\kappa}}_{(1)mn}(\sigma)  \right) \label{81} \\
&&  \kappa_{(0)mn}=e^{2\delta u} {\hat{\kappa}}_{(0)mn}(\sigma), \quad \kappa_{(1)mn}=e^{2\delta u}{\hat{\kappa}}_{(1)mn}(\sigma) \label{82}
\end{eqnarray}
where none of the $``$hatted"
functions of $\sigma$ are specified. \\

\begin{figure}[ht]
    \centering
    \includegraphics[width=8cm]{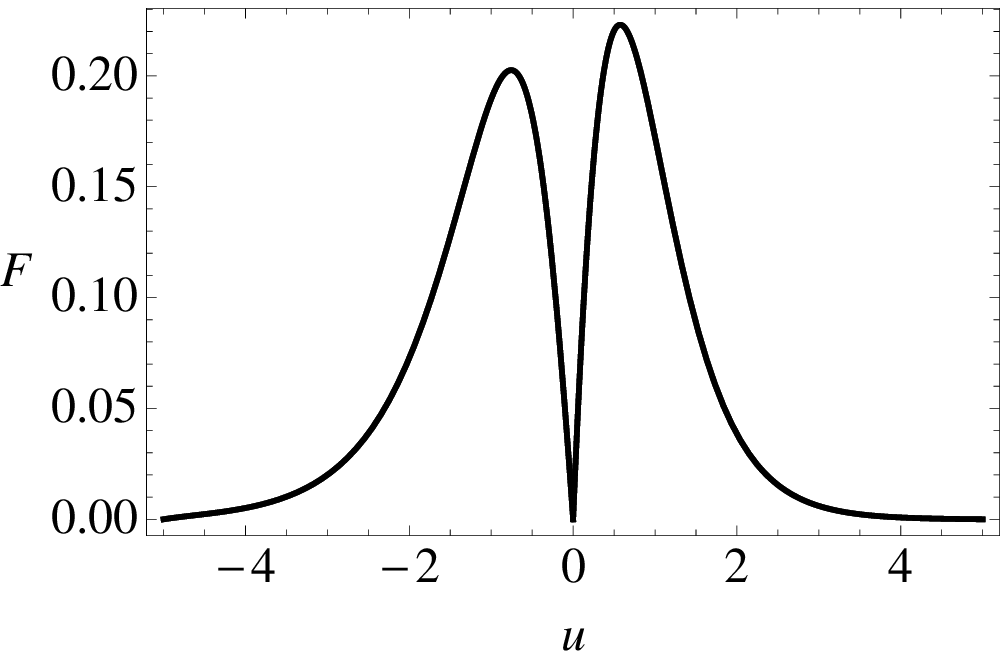}
    \caption{\label{fig:F} Numerical solution for $F$ with $\delta = 0.3$.}
\end{figure}

Now consider the $\phi$ equation of motion. To proceed, one must 
substitute  \eqref{65}, \eqref{66}, \eqref{67}, \eqref{69b} into \eqref{64} noting that it is the trace of $\kappa_{mn}$ defined by
\begin{equation}
\kappa=h^{mn}\kappa_{mn}
\label{lamp2}
\end{equation}
that enters this equation. Expanding 
\begin{equation}
\kappa=\frac{\kappa_{(0)}}{\epsilon}+\kappa_{(1)}+{\cal{O}}(\epsilon) \ ,
\label{lamp3}
\end{equation}
we find using \eqref{81}, \eqref{82} and \eqref{69c} that
\begin{equation}
\kappa_{(0)}=4\delta , \quad \kappa_{(1)}=e^{-2\delta u} {\hat{\kappa}}_{(1)}(\sigma) 
\label{lamp4}
\end{equation}
where ${\hat{\kappa}}_{(1)}={\hat{h}}^{mn}_{(0)}{\hat{\kappa}}_{(1)mn}$. Inserting this along with 
 \eqref{65}, \eqref{66}, \eqref{67} into \eqref{64},
we find to order $\epsilon^{0}$ and  $\epsilon^{1}$ that
\begin{eqnarray}
&&\phi_{(0)}^{\prime\prime}+\kappa_{(0)}\phi_{(0)}^{\prime}-2\phi_{(0)} \left(\phi_{(0)}^{2}-1 \right)+4\delta \left( \phi_{(0)}^{2}-1 \right)= 0 \, \label{83} \\
&& \phi_{(1)}^{\prime\prime} +\kappa_{(0)}\phi_{(1)}^{\prime}+\kappa_{(1)} \phi_{(0)}^{\prime} -2\left( 3\phi_{(0)}^{2}-1 \right)\phi_{(1)} +8\delta \phi_{(0)}\phi_{(1)}=0  \ . \label{84}
\end{eqnarray}
\\
\noindent {\it Order $\epsilon^{0}$:} Using the first expression in \eqref{lamp4}, it is straightforward to show that \eqref{83} has the same solution as in \eqref{53}. That is,
\begin{equation}
\phi_{(0)}=\tanh(u) \ .
\label{85}
\end{equation}
\\
\noindent {\it Order $\epsilon^{1}$:} Using \eqref{lamp4},
equation \eqref{84} becomes
\begin{equation}
\phi_{(1)}^{\prime\prime} +4\delta \phi_{(1)}^{\prime}+e^{-2\delta u}{\hat{\kappa}}_{(1)}(\sigma)\phi_{(0)}^{\prime}-2\left( 3\phi_{(0)}^{2}-1 \right)\phi_{(1)}+8\delta \phi_{(0)}\phi_{(1)}=0 \ .
\label{87}
\end{equation}
Note that as ${\cal{R}} \rightarrow \infty$ and, hence, $\delta \rightarrow 0$, this reverts to the flat space equation \eqref{27}. Let us solve \eqref{87} using the ansatz
\begin{equation}
\phi_{(1)}={\hat{\kappa}}_{(1)}(\sigma)F(u) \ .
\label{88}
\end{equation}
Inserting this into \eqref{87}, the factor ${\hat{\kappa}}_{(1)}$ cancels and one is left with an equation for $F$ given by
\begin{equation}
F^{\prime\prime}+4\delta F^{\prime}+e^{-2\delta u}\phi_{(0)}^{\prime}-2\left(3\phi_{(0)}^{2}-1 \right)F+8\delta\phi_{(0)}F=0 \ .
\label{89}
\end{equation}
Using \eqref{85}, this becomes
\begin{equation}
F^{\prime\prime}-2\left(3~\tanh^{2}(u)-1 \right)F+e^{-2\delta u}{\text{sech}}^{2}(u)+4\delta\left(F^{\prime}+2~\tanh(u)~F \right)=0 \ .
\label{91}
\end{equation}
This equation can be solved independently in each of the separate domains $-\infty <u<0$ and $0<u< \infty$. Corresponding to \eqref{28},\eqref{29}, one must also impose the boundary conditions
\begin{eqnarray}
&& F \rightarrow 0 \quad {\rm as}~ u \rightarrow \pm 0 \ , \label{92} \\
&& F \rightarrow 0 \quad {\rm as}~ u \rightarrow \pm \infty \ .\label{93} 
\end{eqnarray}
Subject to these conditions, there is a unique solution of \eqref{91} which one can solve for numerically. For example, the solution with $\delta=0.3$ is presented in Figure \ref{fig:F}.\\

\noindent We conclude that to this order in $\epsilon$, and restoring the dimensionful parameters, 
\begin{equation}
\Phi=\eta~ \tanh\left(\frac{z}{l}\right)+\eta l~ {\hat{K}}_{(1)}(\sigma) F\left(\frac{z}{l}\right) + {\cal{O}}(\epsilon^{2}) \ .
\label{94}
\end{equation}
As in the flat spacetime case, $\Phi$ is continuous, but not continuously differentiable, across the $z=0$ wall surface.

The worldvolume effective action can now be calculated to any required accuracy in the $\epsilon$ expansion using \eqref{33}, where
\begin{equation}
\hat{\cal{L}}=\int{dz} J {\cal{L}}, \quad   J= \frac{\sqrt{-g}}{\sqrt{-h}|_{u=0}} \ ,
\label{95}
\end{equation}
$g_{mn}$ is the AdS metric \eqref{47} and ${\cal{L}}$ is the original Lagrangian 
density \eqref{49}, \eqref{50} evaluated for the solution of the equation of motion given to order 
$\epsilon$ in \eqref{94}. Taylor expanding $\sqrt{-g}$ around $u=0$ using \eqref{62},\eqref{63} and the fact that 
\begin{equation}
h_{mn}n^{m}n^{n}|_{u=0}=0 \ ,
\label{96}
\end{equation}
one finds that
\begin{equation}
J=1+\epsilon J_{(1)} +\frac{\epsilon^{2}}{2}J_{(2)}+{\cal{O}}(\epsilon^{3}) \ ,
\label{97}
\end{equation}
with
\begin{equation}
J_{(1)}= u \kappa|_{u=0}, \quad J_{(2)}= u^{2}\left(\kappa^{2}-\kappa_{mn} \kappa^{mn}+ 4(\delta/\epsilon)^2\right)|_{u=0} \ .
\label{98}
\end{equation}
Similarly, inserting solution $\Phi$ in \eqref{94} into \eqref{49},\eqref{50} and using the equations of motion \eqref{83},\eqref{87},
it follows that 
\begin{equation}
{\cal{L}}={\cal{L}}_{(0)}+\epsilon {\cal{L}}_{(1)}+\frac{\epsilon^{2}}{2}{\cal{L}}_{(2)}+{\cal{O}}(\epsilon^{3}) \ ,
\label{99}
\end{equation}
where
\begin{equation}
{\cal{L}}_{(i)}={\cal{L}}_{\rm flat (i)}+\delta \Delta_{(i)} \ ,
\label{100}
\end{equation}
the ${\cal{L}}_{\rm flat (i)}$ are of the same functional form as the flat spacetime results in \eqref{38}
and
\begin{eqnarray}
&&\Delta_{(0)}=-\lambda\eta^{4}8\left(\phi_{(0)}-\frac{1}{3}\phi_{(0)}^{3} \right), \quad \Delta_{(1)}=-\lambda\eta^{4}8\phi_{(0)}^{\prime}\phi_{(1)}  \nonumber   \\
&&\qquad \qquad \Delta_{(2)}= -\lambda \eta^{4}8 \left(\phi_{(1)}^{\prime}\phi_{(1)}+\phi_{(0)}^{\prime}\phi_{(2)}  \right) \ .   \label{101} 
\end{eqnarray}
Note that $\delta$ in \eqref{100} arises from the interaction term in the Lagrangian and not from a power series expansion in $\delta$. Hence, $\phi_{(1)}$, $\phi_{(2)}$ in both ${\cal{L}}_{\rm flat (i)}$ and $\Delta_{(i)}$ contain explicit dependence on $\delta$ through, for example, the function $F$. Multiplying \eqref{97} and \eqref{99} gives
\begin{eqnarray}
&&J{\cal{L}} = {\cal{L}}_{\rm flat (0)}\left(1+\frac{\epsilon^{2}}{2}J_{(2)}  \right)+\epsilon {\cal{L}}_{\rm flat (1)}\left(1+\epsilon J_{(1)}  \right)+\frac{\epsilon^{2}}{2}{\cal{L}}_{\rm flat (2)} \nonumber \\
&&\quad +\delta\left(\epsilon(\Delta_{(0)}J_{(1)}+\Delta_{(1)})+\frac{\epsilon^{2}}{2}(\Delta_{(2)}+2\Delta_{(1)}J_{(1)})  \right) +{\cal{O}}(\epsilon^{3}) \ , \label{102} 
\end{eqnarray}
where we have dropped three terms that are odd in $u$ that will vanish when integrated over the transverse coordinate. Note that the 
term proportional to ${\cal{L}}_{\rm flat (0)}$ is identical to the 
${\hat{\cal{L}}}_{(0)}$ term in the  flat spacetime result \eqref{39}, and that the 
$\epsilon^{2}\delta$  terms--which contain $\phi_{(2)}$ in $\Delta_{(2)}$--are subleading. To avoid having to calculate $\phi_{(2)}$, we will, henceforth, work only to lower order. Inserting the solution \eqref{88} for $\phi_{(1)}$ and using the relation
\begin{equation}
{\hat{\kappa}}_{(1)}=\kappa|_{u=0}-\frac{4\delta}{\epsilon} \ ,
\label{103}
\end{equation}
we find that the remaining terms are
\begin{eqnarray}
&&\epsilon {\cal{L}}_{\rm flat (1)}\left(1+\epsilon J_{(1)}  \right)+\frac{\epsilon^{2}}{2}{\cal{L}}_{\rm flat (2)}+\epsilon\delta\left(\Delta_{(0)}J_{(1)}+\Delta_{(1)}\right)=\lambda \eta^{4} \bigg( \epsilon^{2}\kappa|_{u=0}^{2} \phi_{(0)}^{\prime}F \nonumber \\
&&\qquad \qquad -8\epsilon \delta ((\phi_{(0)}-\frac{1}{3}\phi_{(0)}^{3})u+\phi_{(0)}^{\prime}F)\kappa|_{u=0}+16\delta^{2}\phi_{(0)}^{\prime}F \bigg) \label{104}
\end{eqnarray}
plus a total derivative term. We have also ``integrated by parts'' in anticipation of integrating $J{\cal{L}}$ over $u$. 
Putting everything together, \eqref{102} becomes
\begin{eqnarray}
&&\frac{J{\cal{L}}}{\lambda \eta^{4}}=-2\phi_{(0)}^{\prime ~2}-\epsilon^{2}\phi_{(0)}^{\prime~2}u^{2}\left(\kappa^{2}-\kappa_{mn} \kappa^{mn} + 4(\delta/\epsilon)^2 \right)|_{u=0} 
+ \epsilon^{2}\kappa|_{u=0}^{2} \phi_{(0)}^{\prime}F \nonumber \\
&&-8\epsilon \delta ((\phi_{(0)}-\frac{1}{3}\phi_{(0)}^{3})u+\phi_{(0)}^{\prime}F)\kappa|_{u=0}+16\delta^{2}\phi_{(0)}^{\prime}F +{\cal{O}}(\epsilon^{3},\epsilon^{2}\delta,\epsilon \delta^{2},\delta^{3}) \label{105}
\end{eqnarray}
plus a total divergence
in each of the separate domains $-\infty <u<0$ and $0<u<\infty$. To be consistent with dropping the $\epsilon^{2}\delta$ terms above, we only work to quadratic order in the expansion parameters.
Since $F$ in \eqref{105} is multiplied by either $\epsilon^{2}$, $\epsilon \delta$ or $\delta^{2}$, it must be evaluated at order $\delta^{0}$ in \eqref{91}. 
It is important to note from \eqref{lamp4} and \eqref{103} that one can equate
\begin{equation}
\kappa|_{u=0}={\hat{\kappa}}(\sigma) \ ,
\label{105A}
\end{equation}
which we do henceforth.

The previously discussed boundary conditions imply that when integrated over $-\infty<u<\infty$ the contribution of the total divergence term vanishes. Hence, inserting \eqref{105} into \eqref{95} using $z=lu$ and \eqref{54}, one finds
\begin{equation}
\hat{\cal{L}}=\hat{\cal{L}}_{(0)}+\epsilon \hat{\cal{L}}_{(1)}
+\frac{\epsilon^{2}}{2}{\hat{\cal{L}}_{(2)}}+{\cal{O}}(\epsilon^{3},\epsilon^{2}\delta,\epsilon \delta^{2},\delta^{3}) \ ,
\label{106}
\end{equation}
where
\begin{eqnarray}
&&\hat{\cal{L}}_{(0)}=-\frac{\eta^{2}}{l}\left( I_{I} -8\delta^{2} I_{III}\right) \nonumber \\
&&\hat{\cal{L}}_{(1)}=-\frac{4}{l} \eta^{2} \delta \hat{\kappa} \left(I_{\epsilon\delta}+I_{III}\right) \label{107}\\
&&\hat{\cal{L}}_{(2)}=-\frac{\eta^{2}}{l} \left({\hat{\kappa}}^{2}-{\hat{\kappa}}_{mn}{\hat{\kappa}}^{mn} +4 (\delta/\epsilon)^2 \right) I_{II}+\frac{\eta^{2}}{l}{\hat{\kappa}}^{2} I_{III} \nonumber
\end{eqnarray}
and
\begin{eqnarray}
&& I_{I}=\int_{-\infty}^{+\infty}{du~ \phi_{(0)}^{\prime~2}}=\frac{4}{3} \nonumber \\
&& I_{\epsilon \delta}= \int_{-\infty}^{+\infty}{du~u\left(\phi_{(0)}-\frac{1}{3}\phi_{(0)}^{3} \right) }  \label{108}     \\
&&I_{II}=\int_{-\infty}^{+\infty}{du~ u^{2} \phi_{(0)}^{\prime~2}}=\frac{\pi^{2}-6}{9} \nonumber \\
&&I_{III}=\int_{-\infty}^{+\infty}{du~ F~ \phi_{(0)}^{\prime}}=\frac{5}{18} \ . \nonumber
\end{eqnarray}
Note that, unlike $I_{I}$, $I_{II}$ and $I_{III}$, the $I_{\epsilon\delta}$ coefficient primitively diverges like $u^{2}$ as $u \rightarrow \pm\infty$ and, hence, must be carefully treated. Until now, we have loosely taken the range of $u$ to be $-\infty < u < +\infty$. However, this is not strictly correct since the radial Gaussian coordinate is only defined up to the point where the geodesics converge. In our present case, this is either at $L$ or at ${\cal{R}}$, whichever is smallest. That is, the integral has a ``cut-off''.
The effect of this cut-off on the convergent integrals $I_{I}$, $I_{II}$ and $I_{III}$ is negligable and we will, henceforth, ignore it. However,
$I_{\epsilon\delta}$ is now cut-off at $1/\epsilon^{2}$ or $1/\delta^{2}$ respectively, thus rendering it finite. We emphasize that this is completely consistent with taking both the $\sigma^{\mu}$ independent (long wavelength) limit and the flat spacetime limits. In the first case, one takes $\epsilon \rightarrow 0$ holding $\delta$ fixed. Hence, $L>{\cal{R}}$ and $I_{\epsilon\delta} \propto 1/ \delta^{2}$. It follows that all terms in the worldvolume Lagrangian \eqref{106} vanish with the exception of ${\cal{L}}_{(0)}$, and one recovers the lowest order result in AdS spacetime. In the second case, $\delta \rightarrow 0$ holding $\epsilon$ fixed and, hence, ${\cal{R}}>L$ and $I_{\epsilon\delta} \propto 1/ \epsilon^{2}$. Now only the second term in \eqref{106} vanishes, leaving the $\epsilon$ expanded result in flat spacetime given in \eqref{40} of Section \ref{flat}.

Observe that the single-trace extrinsic curvature $\kappa$ appears at each odd order in the Taylor expansion of $\sqrt{-g}$ around $u=0$ ---naively with increasingly singular coefficients. Hence, one might worry that the leading order coefficient $I_{\epsilon \delta}$ given in \eqref{108} is a poor approximation to the the actual value. Again, however, the specific cut-off structure of the $u$-integrals solves this problem. At odd order $n$ in the expansion of $\sqrt{-g}$, there is a primitively divergent integral in the coefficient of $\kappa$. One can show, however, that when appropriately cut-off this becomes
\begin{equation}
  I_{\epsilon \delta}^{(n)} \propto 
  \left\{ 
  \begin{array}{ll}
  1/ \epsilon^{2}~\cdot~\frac{1}{n!}  \left(\delta / \epsilon\right)^{n-1} & \quad \text{if $\delta < \epsilon$}\\ \\
   1 / \delta^{2}~\cdot~\frac{1}{n!}   & \quad \text{if $\epsilon < \delta$} 
   \\
  \end{array} \right.
  \label{108a}
\end{equation} 
Note that for $n=1$, this simply reduces to $I_{\epsilon \delta}$ in either regime. For any odd $n > 1$, it follows from \eqref{108a} that in both regimes $I_{\epsilon \delta}^{(n)} \ll I_{\epsilon \delta}$, increasingly so as $n$ grows. Thus such terms are small compared to the leading term. Computing $I_{\epsilon \delta}$ given in \eqref{108} with the appropriate cut-offs, we find that
\begin{equation}
  I_{\epsilon \delta} = \frac{2}{3} \times
  \left\{ 
  \begin{array}{ll}
  1/ \epsilon^{2} & \quad \text{if $\delta < \epsilon$}\\ 
   1 / \delta^{2} & \quad \text{if $\epsilon < \delta$} 
   \\
  \end{array} \right.
  \label{108b}
\end{equation} 

Rewritten in dimensionful variables, truncating the expansion at order $\epsilon^{2}$ and using
 the Gauss-Codazzi relation 
\begin{equation}
{\hat{R}}^{(4)}={\hat{K}}^{2}-{\hat{K}}_{m}^{n}{\hat{K}}_{n}^{m} - \frac{12}{\mathcal{R}^2} \ ,
\label{109}
\end{equation}
the worldvolume Lagrangian is given by
\begin{equation}
\hat{\cal{L}}=-\frac{4\eta^{2}}{3l}{\cal{M}}_{0}^{4}\left(1+C_{0}{\hat{K}}+ C_{I}{\hat{R}}^{(4)}+C_{II}{\hat{K}}^{2}  \right) \ ,
\label{110}
\end{equation}
where 
\begin{equation}
{\cal{M}}_{0}^{4}=1+6\delta^{2}(I_{II}-I_{III})=1+\frac{2\delta^{2}}{3}\left(\pi^{2}-\frac{17}{4}\right)
\label{111}
\end{equation}
and
\begin{eqnarray}
&& \quad\qquad C_{0}=3\frac{ \left(I_{\epsilon\delta}+I_{III}\right)}{{\cal{M}}_{0}^{4}}l\delta=\left(2 \times
  \left\{ 
  \begin{array}{ll}
  1/ \epsilon^{2}
  \\ 
   1 / \delta^{2} 
   \\
  \end{array} \right. +\frac{5}{6}
  \right) \frac{l\delta}{{\cal{M}}_{0}^{4}}, \label{111A} \\ 
 \nonumber \\  
&&C_{I}=\frac{ I_{II}}{{\cal{M}}_{0}^{4}}\frac{3l^{2}}{8}=\left(\frac{\pi^{2}-6}{24} \right)\frac{l^{2}}{{\cal{M}}_{0}^{4}}, \quad
C_{II}=-\frac{ I_{III} }{{\cal{M}}_{0}^{4}}\frac{3l^{2}}{8}=-\frac{5}{48}\frac{l^{2}}{{\cal{M}}_{0}^{4}} \ . \nonumber
\end{eqnarray}
As discussed above, in the limit ${\cal{R}}\rightarrow \infty$ and, hence, $\delta \rightarrow 0$, Lagrangian \eqref{110} becomes 

\begin{figure}[H]
    \centering
    \begin{tabular}{cc}
    \includegraphics[width=6cm]{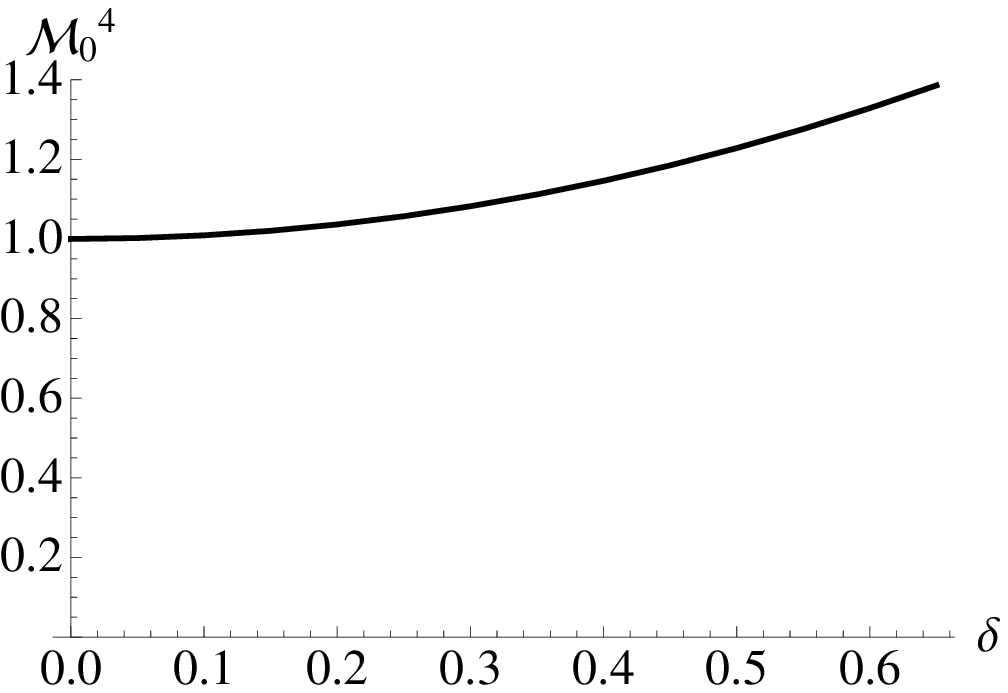}{(A)} & \includegraphics[width=6cm]{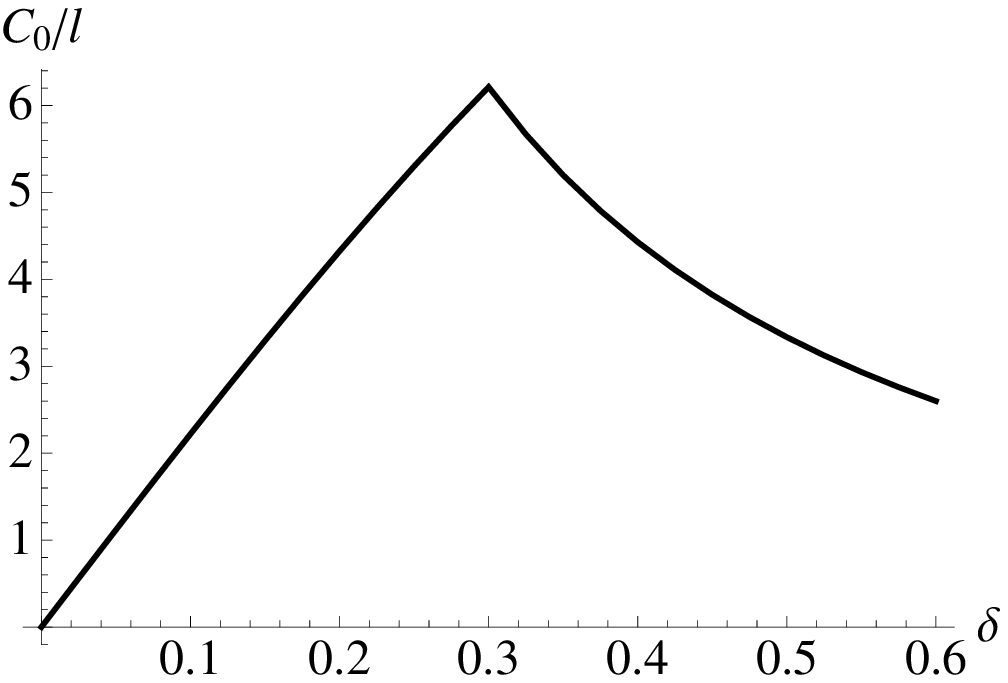}(B) \\
    \includegraphics[width=6cm]{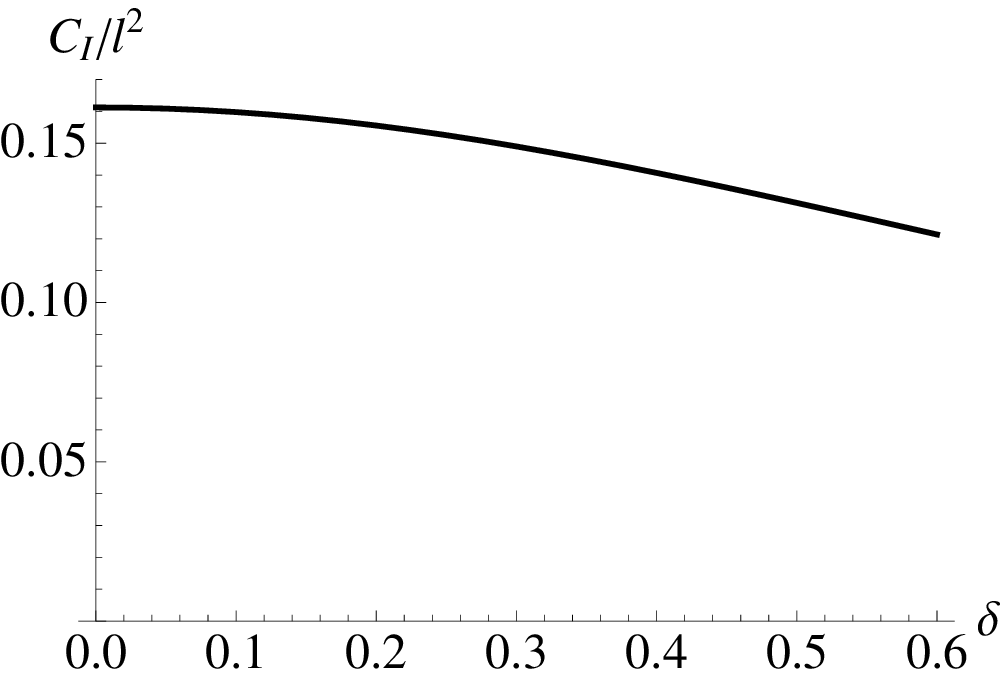}(C) & \includegraphics[width=6cm]{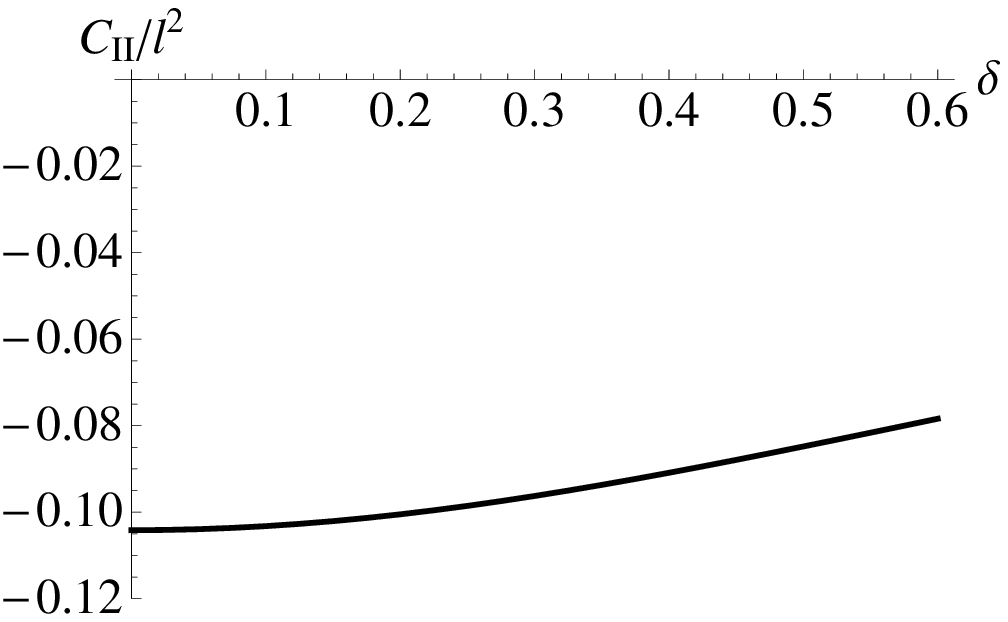}(D) \\
    \end{tabular}      
   \caption{\label{fig:C} Numerical calculation of ${\cal{M}}_{0}^{4}$, 
    $C_0 / l$, $C_{I} / l^{2}$ and $C_{II} / l^{2}$ as functions of $\delta$. Of the four coefficients, only $C_{0} /l $ depends on $I_{\epsilon \delta}$ and, hence, on the value of the cut-off ratio $\frac{\cal{R}}{L}=\frac{\epsilon}{\delta}$. Therefore, to evaluate $C_{0} / l$ we must specify a value for $\epsilon$. In Figure 2(B), we choose $\epsilon = 0.3$. Note that $C_0 / l$ is defined piecewise and changes behavior at $\delta \sim \epsilon$.}
\end{figure}

\noindent the flat spacetime Lagrangian given in \eqref{43},\eqref{45}. Explicit values of the overall coefficient ${\cal{M}}_{0}^{4}$,
as well as the three coupling parameters $C_{0} / l$, $C_{I} /l^{2}$ and $C_{II}/ l^{2}$, can be plotted numerically as functions of $\delta$. These are shown as graphs (A),(B),(C)and (D) respectively in  Figure \ref{fig:C}. Note that $C_0$ has the correct limiting value $C_0 \to 0$ as $\delta \to 0$.

\subsection*{A Simplified ``Flat'' Metric Approach \label{newflat}}

The non-vanishing curvature \eqref{48} of AdS space considerably complicates the previous analysis; specifically,
leading to a non-zero value for $K_{(0)mn}$ which then propagates through the calculation. However, recall that the metric \eqref{47} is conformally flat. This is manifest in the new coordinate $u^{\prime}=z^{\prime}/l$ defined by
\begin{equation}
u^{\prime}=\frac{1}{\delta}\left(1-e^{-\delta u} \right) \ , \quad -\infty< u^{\prime}< \frac{1}{\delta} 
\label{112}
\end{equation}
with respect to which metric \eqref{47} becomes
\begin{equation}
ds^{2}=e^{2\delta u} \left(dx^{\mu}dx^{\nu}\eta_{\mu\nu}+dz^{\prime 2}  \right) \ .
\label{112A}
\end{equation}
This motivates us to repeat the AdS analysis in terms of the rescaled flat $``$metric"
\begin{equation}
{\tilde{g}}_{mn}=e^{-2\delta u}g_{mn} ~\Rightarrow~ {\tilde{h}}_{mn}=e^{-2\delta u}h_{mn} \ ,
\label{113}
\end{equation}
since we expect ${\tilde{K}}_{(0)mn}$ to vanish.

Inserting \eqref{113} into \eqref{56} we find 
\begin{equation}
{\cal{L}}_{\tilde{n}}{\tilde{h}}_{mn}=2{\tilde{K}}_{mn} \ ,
\label{114}
\end{equation}
where
\begin{equation}
{\tilde{K}}_{mn}=\left(K_{mn}-\frac{1}{{\cal{R}}}h_{mn} \right)e^{-\delta u} \ .
\label{115}
\end{equation}
Similarly, putting \eqref{115} into \eqref{59} gives
\begin{equation}
{\cal{L}}_{\tilde{n}}{\tilde{K}}_{mn}={\tilde{K}}_{mp}{\tilde{K}}^{p}_{n}-\frac{1}{{\cal{R}}}{\tilde{K}}_{mn}e^{\delta u} \ .
\label{116}
\end{equation}
We note that the curvature term on the right-hand side of \eqref{59} cancels when going to the flat metric. However, the conformal rescaling induces the term proportional to $1/{\cal{R}}$ in \eqref{116}.
In addition to \eqref{114} and \eqref{116}, one must also re-express the equation of motion 
\eqref{60} with respect to the rescaled flat metric. We find that
\begin{eqnarray}
&&{\cal{L}}_{\tilde{n}}^{2}\Phi+{\tilde{K}}{\cal{L}}_{{\tilde{n}}}\Phi+\frac{3}{{\cal{R}}}e^{\delta u}{\cal{L}}_{{\tilde{n}}} \Phi+D^{m}D_{m}\Phi \nonumber \\
&&\qquad \qquad  -e^{2\delta u}\left( 4\lambda \Phi(\Phi^{2}-\eta^{2})+\frac{4\sqrt{2\lambda}}{{\cal{R}}}\left( \eta^{2}-\Phi^{2} \right)\right) =0 \ .
\label{117}
\end{eqnarray}
Going to dimensionless variables using \eqref{61},
equations \eqref{114},\eqref{116} and \eqref{117} become
\begin{eqnarray}
&&{\tilde{h}}_{mn}^{\prime}=2\epsilon {\tilde{\kappa}}_{mn} \ , \label{118} \\
&&{\tilde{\kappa}}_{mn}^{\prime}=\epsilon {\tilde{\kappa}}_{mp}{\tilde{\kappa}}_{qn}{\tilde{h}}^{pq} -\delta {\tilde{\kappa}}_{mn}e^{\delta u} \ ,  \label{119} \\
&&\phi^{\prime\prime}+\left(\epsilon {\tilde{\kappa}}+3\delta e^{\delta u}\right) \phi^{\prime}-2 e^{2\delta u}(\phi-2\delta)(\phi^{2}-1)+\epsilon^{2}D^{m}D_{m}\phi=0 \label{120}
\end{eqnarray}
where, now, $^{\prime}=\frac{\partial}{\partial u^{\prime}}$.

These equations are solved using the $\epsilon$-expansions in \eqref{65},\eqref{66},
\eqref{67} and \eqref{69b}, now expressed in terms of ${\tilde{h}}_{mn}$ and ${\tilde{\kappa}}_{mn}$ quantities.
First consider the ${\tilde{h}}_{mn}$ equation \eqref{118}. We find to  order $\epsilon^{0}$ 
and $\epsilon^{1}$ that
\begin{eqnarray}
&&{\tilde{h}}^{\prime}_{(0)mn}=2{\tilde{\kappa}}_{(0)mn}\ , \label{121}  \\
&&{\tilde{h}}^{\prime}_{(1)mn}=2{\tilde{\kappa}}_{(1)mn} \label{122} 
\end{eqnarray}
respectively. Similarly, expanding the ${\tilde{\kappa}}_{mn}$ equation \eqref{119} we find to  order $\epsilon^{0}$ and  $\epsilon^{1}$ that
\begin{eqnarray}
&& {\tilde{\kappa}}^{\prime}_{(0)mn}={\tilde{\kappa}}_{(0)mp}{\tilde{\kappa}}_{(0)qn}{\tilde{h}}_{(0)}^{pq} -\delta e^{\delta u}{\tilde{\kappa}}_{(0)mn} \ , \label{123} \\
&& {\tilde{\kappa}}^{\prime}_{(1)mn}={\tilde{\kappa}}_{(0)mp}{\tilde{\kappa}}_{(1)qn}{\tilde{h}}_{(0)}^{pq} +{\tilde{ \kappa}}_{(1)mp}{\tilde{\kappa}}_{(0)qn}{\tilde{h}}_{(0)}^{pq} \label{124} \\
&&\qquad  \qquad+{\tilde{\kappa}}_{(0)mp} {\tilde{\kappa}}_{(0)qn} {\tilde{h}}_{(1)}^{pq} -\delta e^{\delta u}{\tilde{\kappa}}_{(1)mn}  \ . \nonumber
\end{eqnarray}
Before continuing to the equation of motion, let us solve \eqref{121},\eqref{122} and \eqref{123},
\eqref{124}. \\

\noindent {\it Order $\epsilon^{0}$:} Since at this order ${\tilde{n}}_{m}={\tilde{n}}_{5}$, it follows from \eqref{112A},\eqref{113} that 
\begin{equation}
{\tilde{h}}_{(0)mn}={\hat{\tilde{h}}}_{(0)mn}(\sigma) 
\label{125}
\end{equation}
with $\hat{\tilde{h}}_{(0)mn}$ unspecified.
Hence, the ${\tilde{h}}_{(0)mn}$ equation \eqref{121} gives
\begin{equation}
{\tilde{\kappa}}_{(0)mn}=0 \ ,
\label{126}
\end{equation}
as expected.
It follows immediately that the ${\tilde{\kappa}}_{(0)mn}$ equation \eqref{123} is trivially satisfied.\\

\noindent {\it Order $\epsilon^{1}$:} Substituting \eqref{126} into the order $\epsilon^{1}$ ${\tilde{\kappa}}_{mn}$ equation \eqref{124}, recalling that $^{\prime}=\frac{\partial}{\partial u^{\prime}}$ and using \eqref{112}, we find
\begin{equation}
{\tilde{\kappa}}_{(1)mn}=e^{-\delta u}{\hat{\tilde{\kappa}}}_{(1)mn}(\sigma) 
\label{127}
\end{equation}
with ${\hat{\tilde{\kappa}}}_{(1)}$ an arbitrary function of $\sigma^{\mu}$-coordinates only. Finally, inserting this expression into the ${\tilde{h}}_{(1)mn}$ equation \eqref{122} gives
\begin{equation}
{\tilde{h}}_{(1)mn}=-\frac{1}{\delta}\left(e^{-\delta u}{\tilde{\kappa}}_{(1)mn}-{\hat{\tilde{\kappa}}}_{(1)mn}(\sigma) \right) \ .
\label{128}
\end{equation}
\\
 Using the relation \eqref{115}, these results are easily compared against the $g_{mn}$ metric results summarized in \eqref{81},\eqref{82}. Identifying the arbitrary functions
\begin{equation}
{\hat{\tilde{h}}}_{(0)mn}(\sigma)=\frac{1}{\delta}{\hat{\kappa}}_{(0)mn}(\sigma), \quad {\hat{\tilde{\kappa}}}_{(1)mn}(\sigma)={\hat{\kappa}}_{(1)mn}(\sigma)
\label{129}
\end{equation}
we find exact agreement.

Now consider the $\epsilon$-expansion of the $\phi$ equation of motion \eqref{120}. The trace
\begin{equation}
{\tilde{\kappa}}={\tilde{h}}^{mn}{\tilde{\kappa}}_{mn}
\label{130}
\end{equation}
enters this equation. Expanding 
\begin{equation}
{\tilde{\kappa}}=\frac{{\tilde{\kappa}}_{(0)}}{\epsilon}+{\tilde{\kappa}}_{(1)}+{\cal{O}}(\epsilon) \ ,
\label{131}
\end{equation}
we find using \eqref{125}-\eqref{128} and \eqref{129} that
\begin{equation}
{\tilde{\kappa}}_{(0)}=0 , \quad {\tilde{\kappa}}_{(1)}=e^{-\delta u} {\hat{\kappa}}_{(1)}(\sigma) 
\label{132}
\end{equation}
where ${\hat{\kappa}}_{(1)}={\hat{h}}^{mn}_{(0)}{\hat{\kappa}}_{(1)mn}$. Note from \eqref{113}
and \eqref{115} that 
\begin{equation}
{\tilde{\kappa}}=\left(\kappa-\frac{4\delta}{\epsilon} \right)e^{\delta u} \ .
\label{132A}
\end{equation}
It follows that \eqref{132} is completely consistent with the $g_{mn}$ metric results in \eqref{lamp4}. 
Inserting \eqref{132} along with the $\epsilon$-expansions of $\phi$, ${\tilde{h}}_{mn}$ and ${\tilde{\kappa}}_{mn}$ into \eqref{120},
we find to order $\epsilon^{0}$ and  $\epsilon^{1}$ that
\begin{eqnarray}
&&\phi_{(0)}^{\prime\prime}+3\delta e^{\delta u}\phi_{(0)}^{\prime}-2e^{2\delta u}\left(\phi_{(0)}^{2}-1 \right)\left( \phi_{(0)}-2\delta \right)= 0 \, \label{133} \\
&& \phi_{(1)}^{\prime\prime} +3\delta e^{\delta u}\phi_{(1)}^{\prime}+e^{-\delta u}{\hat{\kappa}}_{(1)}(\sigma) \phi_{(0)}^{\prime} \label{134} \\
&& \qquad ~\quad+e^{2\delta u} \left(-2\left( 3\phi_{(0)}^{2}-1 \right) +8\delta \phi_{(0)}\right)\phi_{(1)}=0  \ . \nonumber
\end{eqnarray}
\\

\noindent {\it Order $\epsilon^{0}$:} It is straightforward to show that \eqref{133} has the same solution as equation \eqref{83}, although expressed in the $u^{\prime}$ coordinate. That is,
\begin{equation}
\phi_{(0)}=\tanh(u), \quad u=-\frac{1}{\delta}\ln\left(1-\delta u^{\prime} \right) 
\label{135}
\end{equation}
where we have inverted expression \eqref{112}.
\\

\begin{figure}[h]
    \centering
    \includegraphics[width=8cm]{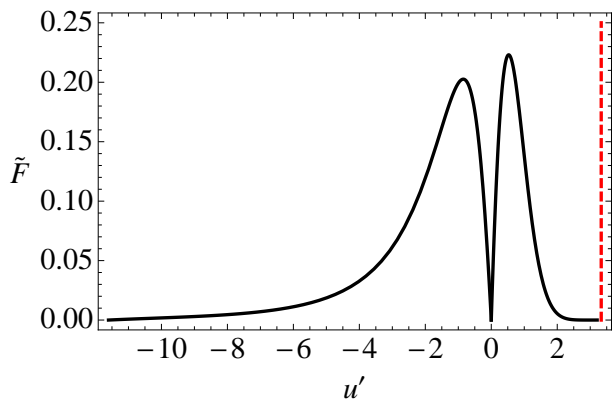}
    \caption{\label{fig:F2} Numerical solution for ${\tilde{F}}(u^{\prime})$ with $\delta = 0.3$.}
\end{figure}

\noindent {\it Order $\epsilon^{1}$:} Let us solve \eqref{134} using the ansatz
\begin{equation}
\phi_{(1)}={\hat{\kappa}}_{(1)}(\sigma){\tilde{F}}(u^{\prime}) \ .
\label{136}
\end{equation}
Inserting this into \eqref{134}, the factor ${\hat{\kappa}}_{(1)}$ cancels and one is left with an equation for ${\tilde{F}}$ given by
\begin{equation}
{\tilde{F}}^{\prime\prime}+3\delta e^{\delta u}{\tilde{F}}^{\prime}+e^{-\delta u}\phi_{(0)}^{\prime}+e^{2\delta u} \left(-2\left(3\phi_{(0)}^{2}-1 \right)+8\delta\phi_{(0)}\right){\tilde{F}}=0 
\label{137}
\end{equation}
with $\phi_{(0)}(u^{\prime})$ in \eqref{135}.
This equation can be solved independently in each of the separate domains $-\infty <u^{\prime}<0$ and $0<u^{\prime}< \frac{1}{\delta}$. Corresponding to \eqref{28},\eqref{29}, one must also impose the boundary conditions
\begin{eqnarray}
&& {\tilde{F}} \rightarrow 0 \quad {\rm as}~ u^{\prime} \rightarrow \pm 0 \ , \label{138} \\
&& {\tilde{F}} \rightarrow 0 \quad {\rm as}~ u^{\prime} \rightarrow -\infty,~\frac{1}{\delta} \ .
\label{139} 
\end{eqnarray}
Subject to these conditions, there is a unique solution of \eqref{137} which one can solve for numerically. For example, the solution with $\delta=0.3$ is presented in Figure \ref{fig:F2}.\\
Note that since the same factor ${\hat{\kappa}}_{(1)}(\sigma)$ enters both \eqref{88} and 
\eqref{136}, it follows that when re-expressed in terms of the coordinate $u$ using \eqref{112} one must find
\begin{equation}
{\tilde{F}}(u^{\prime})=F(u) \ .
\label{140}
\end{equation}
Comparing the ${\tilde{F}}$ and $F$ results for $\delta=0.3$ given in Figures 1 and  3 respectively,
we find complete agreement. We conclude that to this order in $\epsilon$, and restoring the dimensionful parameters, one again finds
\begin{equation}
\Phi=\eta~ \tanh\left(\frac{z}{l}\right)+\eta l~ {\hat{K}}_{(1)}(\sigma) F\left(\frac{z}{l}\right) + {\cal{O}}(\epsilon^{2}) \ .
\label{141}
\end{equation}
where
\begin{equation}
\frac{z}{l}=-\frac{1}{\delta}\ln\left(1-\delta \frac{z^{\prime}}{l} \right) 
\label{142}
\end{equation}
and $\Phi$ is continuous, but not continuously differentiable, across the $z^{\prime}=0$ wall surface.

The worldvolume action can be calculated with respect to the rescaled ${\tilde{g}}_{mn}$ metric to any desired degree of accuracy in the $\epsilon$ expansion. It follows from our previous discussion that
\begin{equation}
\hat{\cal{L}}=\int{dz^{\prime} \tilde{J} \tilde{\cal{L}}}, \quad \tilde{J}=\frac{\sqrt{-\tilde{g}}}
{\sqrt{-\tilde{h}}|_{u^{\prime}=0}  }
\label{142a}
\end{equation}
where $\tilde{\cal{L}}$ is the Lagrange density which, when varied with respect to $\Phi$, gives the equation of motion \eqref{117}. It is found to be
\begin{equation}
 \tilde{\cal{L}}=e^{5\delta u}{\cal{L}}
\label{142b}
\end{equation}
with ${\cal{L}}$ given by \eqref{49},\eqref{50} written in the $z^{\prime}$ coordinate. In \eqref{142a}, $\tilde{\cal{L}}$ is to be evaluated for the solution of the equation of motion given, to order 
$\epsilon$, in \eqref{141},\eqref{142}. Taylor expanding $\sqrt{-\tilde{g}}$ around $u^{\prime}=0$, we find 
\begin{equation}
\tilde{J}=1+\epsilon \tilde{J}_{(1)} +\frac{\epsilon^{2}}{2}\tilde{J}_{(2)}+{\cal{O}}(\epsilon^{3}) \ ,
\label{142c}
\end{equation}
with
\begin{equation}
\tilde{J}_{(1)}= u^{\prime} \tilde{\kappa}|_{u^{\prime}=0}, \quad \tilde{J}_{(2)}= u^{\prime 2}\left(\tilde{\kappa}^{2}-\tilde{\kappa}_{mn} \tilde{\kappa}^{mn}- (\delta/ \epsilon)\tilde{\kappa}\right)|_{u^{\prime}=0} \ .
\label{142d}
\end{equation}

To continue, we note that ${\cal{L}}$ is coordinate invariant and that $\Phi$ is most conveniently expressed as a function of the coordinate $z$. It follows that the $\epsilon$ expansion of $\cal{L}$ is most easily presented by going back to the coordinate $z$ using \eqref{112}. Noting that 
\begin{equation}
u^{\prime}=u-\frac{\delta}{2}u^{2}+\dots , \qquad dz^{\prime}e^{\delta u}=dz  
\label{142dd}
\end{equation}
we can write action \eqref{142a},\eqref{142b} as
\begin{equation}
\hat{\cal{L}}=\int{dz e^{4\delta u}\tilde{J} {\cal{L}}} \ ,
\label{142ddd}
\end{equation}
where the coefficients of the expansion \eqref{142c} with respect to the $u$ coordinate are
\begin{equation}
\tilde{J}_{(1)}= u \tilde{\kappa}|_{u=0}, \quad \tilde{J}_{(2)}= u^{2}\left(\tilde{\kappa}^{2}-\tilde{\kappa}_{mn} \tilde{\kappa}^{mn}- 2(\delta/\epsilon)\tilde{\kappa}\right)|_{u=0} \ .
\label{142dddd}
\end{equation}
Similarly, inserting the solution $\Phi$ into \eqref{142b} and using the equations of motion \eqref{133},\eqref{134} it follows that
\begin{equation}
{\cal{L}}={\cal{L}}_{(0)}+\epsilon {\cal{L}}_{(1)}+\frac{\epsilon^{2}}{2}{\cal{L}}_{(2)}+{\cal{O}}(\epsilon^{3}) \ ,
\label{142e}
\end{equation}
where
\begin{equation}
{\cal{L}}_{(i)}={\cal{L}}_{\rm flat (i)}+\delta \Delta_{(i)} \ .
\label{142f}
\end{equation}
${\cal{L}}_{\rm flat (i)}$ are the flat spacetime results in \eqref{38}, $\Delta_{(i)} $ were given in \eqref{101} and all quantities are functions of $z$. Finally, one must expand the exponential
\begin{equation}
e^{4\delta u} = 1 +4\delta u +8\delta^{2}u^{2}+ {\cal{O}}((\delta u)^{3}) \ .
\label{142ff}
\end{equation}

Let us proceed order by order in the $\delta u$ expansion, keeping only those terms that are even under $u\rightarrow -u$. The $1$-term is of the same form as given in \eqref{102} of the proceeding section, now, however,  with the $J_{(i)}$ replaced by $\tilde{J}_{(1)}$,$\tilde{J}_{(2)}$ in \eqref{142dddd}. Furthermore, in the present case $\hat{\kappa}_{(1)}={\tilde{\kappa}}|_{u=0
}$. Hence, the constant term on the right side of \eqref{103} no longer appears.
We find that the $1$-term in the $\delta u$ expansion is given by
\begin{eqnarray}
&&\frac{\tilde{J}{\cal{L}}}{\lambda \eta^{4}}\big|_{1}=-2\phi_{(0)}^{\prime ~2}-\epsilon^{2}\phi_{(0)}^{\prime~2}u^{2}\left(\tilde{\kappa}^{2}-\tilde{\kappa}_{mn} \tilde{\kappa}^{mn} -2(\delta/\epsilon)\tilde{\kappa} \right)|_{u=0} + \epsilon^{2}\tilde{\kappa}|_{u=0}^{2} \phi_{(0)}^{\prime}F \nonumber \\
&&\qquad \qquad -8\epsilon \delta ((\phi_{(0)}-\frac{1}{3}\phi_{(0)}^{3})u+\phi_{(0)}^{\prime}F)\tilde{\kappa}|_{u=0}+{\cal{O}}(\epsilon^{3},\epsilon^{2}\delta,\epsilon \delta^{2},\delta^{3}) \ . \label{142g}
\end{eqnarray}
Next consider the $\delta u$-term. Using \eqref{38},\eqref{101} and \eqref{142dddd}, to the order we are working 
\begin{equation}
\frac{\tilde{J}{\cal{L}}}{\lambda \eta^{4}}\big|_{\delta u}=-32\delta^{2}(\phi_{(0)}-\frac{1}{3}\phi_{(0)}^{3})u-8\epsilon\delta \left(\phi_{(0)}^{\prime ~2}u^{2}-\phi_{(0)}^{\prime}F\right){\tilde{\kappa}}|_{u=0} \ .
\label{142gg}
\end{equation}
Finally, the last term one need consider is the $(\delta u)^{2}$-term.  We find that
\begin{equation}
\frac{\tilde{J}{\cal{L}}}{\lambda \eta^{4}}\big|_{(\delta u)^{2}}=-16\delta^{2}\phi_{(0)}^{\prime ~2}u^{2} \ .
\label{142h}
\end{equation}
Adding these together, inserting them into \eqref{142ddd}, using $z=lu$, \eqref{54} and denoting
\begin{equation}
{\tilde{\kappa}}|_{u=0}=\hat{\tilde{\kappa}}(\sigma) \ ,
\label{ink}
\end{equation}
we find that
\begin{equation}
\hat{\cal{L}}={\hat{\tilde{\cal{L}}}}_{(0)}+\epsilon {\hat{\tilde{\cal{L}}}}_{(1)}+\frac{\epsilon^{2}}{2}{\hat{\tilde{\cal{L}}}}_{(2)}+{\cal{O}}(\epsilon^{3},\epsilon^{2}\delta,\epsilon \delta^{2},\delta^{3}) \ ,
\label{142hh}
\end{equation}
where
\begin{eqnarray}
&&{\hat{\tilde{\cal{L}}}}_{(0)}=-\frac{\eta^{2}}{l}\left( I_{I} +16\delta^{2} I_{\epsilon\delta}+8\delta^{2}I_{II}\right) \nonumber \\
&&{\hat{\tilde{\cal{L}}}}_{(1)}=-\frac{4\eta^{2} }{l} \delta \hat{\tilde{\kappa}} \left(I_{\epsilon\delta}+I_{II}\right) \label{142i}\\
&&{\hat{\tilde{\cal{L}}}}_{(2)}=-\frac{\eta^{2}}{l} \left({\hat{\tilde{\kappa}}}^{2}-{\hat{\tilde{\kappa}}}_{mn}{\hat{\tilde{\kappa}}}^{mn} -2 (\delta/\epsilon)\hat{\tilde{\kappa}} \right) I_{II}+\frac{\eta^{2}}{l}{\hat{\tilde{\kappa}}}^{2} I_{III} \nonumber
\end{eqnarray}
with the $I$-coefficients given in \eqref{108}. 

Rewritten in dimensionful variables and truncating the expansion at order $\epsilon^{2}$, the worldvolume Lagrangian is given by
\begin{eqnarray}
&&\frac{\hat{\cal{L}}}{-\eta^{2}/l}=\left( I_{I} +8\delta^{2}(2 I_{\epsilon\delta}+\delta^{2}I_{II})\right)+l\delta\left(4I_{\epsilon\delta}+3I_{II} \right) \hat{\tilde{K}} \nonumber \\
&& \qquad\qquad +\frac{l^{2}}{2}(I_{II}) {\hat{\tilde{R}}}^{(4)} -\frac{l^{2}}{2}(I_{III}) {\hat{\tilde{K}}}^{2} \ ,
\label{142ii}
\end{eqnarray}
where the Gauss-Codazzi relation
\begin{equation}
{\hat{\tilde{R}}}^{(4)}={\hat{\tilde{K}}}^{2}-{\hat{\tilde{K}}}_{m}^{n}{\hat{\tilde{K}}}_{n}^{m} 
\label{142j}
\end{equation}
has been used. Note that the $ \hat{\tilde{K}}$, ${\hat{\tilde{R}}}^{(4)}$and
${\hat{\tilde{K}}}^{2}$ terms all appear in \eqref{142ii}, although with differing coefficients then in the previous section. What is the relationship of this expression to the worldvolume Lagrangian \eqref{109} evaluated with respect to the $g_{mn}$ metric? Note from \eqref{115} that at $u=0$
\begin{equation}
\hat{\tilde{K}}_{mn}=\hat{K}_{mn}-\frac{1}{{\cal{R}}}h_{mn} \ .
\label{142k}
\end{equation}
Inserting this into \eqref{142ii}, a careful calculation reveals that it is identical to \eqref{109}, as it must be.

\section{The Worldvolume Action and Galileons \label{worldvolume}}

It follows from the previous section that the worldvolume action of a $d=4$ kink domain wall embedded in $d=5$ anti-deSitter spacetime, to second order in the $\epsilon$-expansion, is 
\begin{equation}
S_{4}=\int_{M^{4}} {d^{4}\sigma \sqrt{-h}|_{u=0}{\hat{\mathcal{L}}}}
\label{143}
\end{equation}
where
\begin{equation}
{\hat{\mathcal{L}}}=-\frac{4\eta^{2}}{3l}{\mathcal{M}}_{0}^{4} \left(1+C_{0}{\hat{K}}+ C_{I}{\hat{R}}^{(4)}+C_{II}{\hat{K}}^{2}  \right) \ .
\label{144}
\end{equation}
The coefficients ${\cal{M}}_{0}^{4}$ 
and $C_{0}$, $C_{I}$, $C_{II}$ are given in \eqref{111} and \eqref{111A} respectively. Note that $C_{0}$ is proportional to the kink thickness $l$, whereas both $C_{I}$ and $C_{II}$ are proportional to $l^{2}$---corresponding in dimensionless variables to $\epsilon$ and $\epsilon^{2}$ respectively.   
Recall that the embedding of a $d=4$ worldvolume in a $d=5$ bulk space is specified by five worldvolume scalar fields $X^{m}(\sigma)$, $m=0,1,\dots,4$. Choosing the gauge where 
\begin{equation}
X^{\mu}=\sigma^{\mu}, ~\mu=0,1,2,3 ~~ , \quad X^{4} \equiv \pi(\sigma) 
\label{146}
\end{equation}
we find that
\begin{eqnarray}
&&\sqrt{-h}|_{u=0}=e^{4\pi/\mathcal{R}} \sqrt{1+ e^{-2\pi/\mathcal{R}}(\partial \pi)^2} \ ,  \label{147} \\
&&{\hat{K}}=  -e^{-2\pi/\mathcal{R}}\tilde{\gamma}
 \left(-\square \pi+\tilde{\gamma}^2 e^{-2\pi/\mathcal{R}}[\phi]+\frac {\tilde{\gamma}^2}{\mathcal{R}} (\partial \pi)^2+\frac{4}{\mathcal{R}} e^{2\pi/\mathcal{R}} \right), \label{148} \\
&&{\hat{R}}^{(4)}  = \tilde \gamma^4 e^{-4\pi/\mathcal{R}}\Bigg[
\tilde{\gamma}^{-2}\left((\square\pi)^2-(\partial_\mu\partial_\nu \pi)^2\right)+2 e^{-2\pi/\mathcal{R}}\bigg([\phi^2] \nonumber \\
&&\qquad \quad -(\square\pi)[\phi]\bigg) \label{149}-\frac{6}{\mathcal{R}^2}e^{2\pi/\mathcal{R}}(\partial \pi)^2 \left(1+2 e^{-2\pi/\mathcal{R}}(\partial \pi)^2\right) \label{149}\\
&&\qquad \quad  +\frac{8}{\mathcal{R}} [\phi] 
-\frac{2}{\mathcal{R}} e^{2\pi/\mathcal{R}}\left(3+4e^{-2\pi/\mathcal{R}}(\partial \pi)^2\right)\square\pi
\Bigg] \notag
\end{eqnarray}
and
\begin{equation}
{\hat{K}}^{2}= e^{-4\pi/\mathcal{R}}\tilde{\gamma}^2
 \left(-\square \pi+\tilde{\gamma}^2 e^{-2\pi/\mathcal{R}}[\phi]+\frac {\tilde{\gamma}^2}{\mathcal{R}} (\partial \pi)^2+\frac{4}{\mathcal{R}} e^{2\pi/\mathcal{R}} \right)^2
\label{150}
\end{equation}
where $\square = \eta^{\mu\nu}\partial_\mu \partial_\nu$, $\tilde{\gamma} = 1/\sqrt{1+ e^{-2\pi/\mathcal{R}}(\partial \pi)^2}$, $[\phi] = \partial^\mu\pi \partial_\mu\partial^\nu  \pi \partial_\nu \pi$, $[\phi^2] = \partial_\mu \pi \partial^\mu \partial_\nu \pi \partial^\nu \partial_\lambda \pi\partial^\lambda\pi $ and indices are raised and lowered with respect to the flat metric $\eta^{\mu\nu}$.

Note that $\sqrt{-h}|_{u=0}$, $\sqrt{-h}|_{u=0}{\hat{K}}$ and $\sqrt{-h}|_{u=0}{\hat{R}}^{(4)}$ correspond to the $L_{2}$, $L_{3}$ and $L_{4}$ DBI conformal Galileons introduced and calculated in \cite{deRham:2010eu}, and lead to second order equations of motion for $\pi$. The $L_{5}$ Galileon is a higher-dimension operator---arising at cubic order in the $\epsilon$-expansion---and, hence, does not appear here.  
However, ${\hat{K}}^{2}$ given in \eqref{150} is {\it not a Galileon}.  It will lead to fourth-order equations of motion, but any ghosts or pathologies associated with this operator will not appear in the range of validity of the $\epsilon$-expansion, since it was derived from a ghost free theory.
Furthermore, carefully analyzing the above expressions we find that there is no region of $\pi$ or momentum space for which non-Galileon terms such as the $\hat{K^{2}}$ term is sub-dominant to the Galileon terms.  Thus, in any situation in which the Galileons are important relative to the kinetic term and their non-linearities are doing something interesting, the non-Galileon terms are important as well, and the entire series expansion we are computing is breaking down.   We conclude, to the order we have calculated, that although Galileon terms appear in the explicitly computed worldvolume action, there is an additional non-Galileon term which can not be neglected, and that to justify stopping at some order in the expansion, all these terms must be subdominant to the kinetic term.

Having specified this, it is interesting to note that, by an appropriate redefinition of the $\pi$ field, the effective action \eqref{143} can be written so that only Galileon operators appear. We prove this in Appendix A to cubic order in the $\epsilon$-expansion---one order higher than the results of this paper. Be this as it may, the non-Galileon  ${\hat{K}}^{2}$ term, although removed from ${\hat{\mathcal{L}}}$, is now manifest in the field redefinition.  If one is interested in computing quantities which are independent of field redefinitions, such as scattering amplitudes, then it suffices to use only the Galileon interactions, and the non-Galileon terms do not affect these quantities.   However, if one is interested in quantities that do depend on the definition of $\pi$, such as computing the  physical location of the brane, the presence of the non-Galileon terms matters.

\section*{Appendix A}

\subsection*{Field Redefinitions:}

First a general argument about field redefinitions.  Let $\mathcal{L}$ be a Lagrangian density for fields $\phi^i$ which is a formal series in some parameter $\lambda$, 
\begin{equation}\mathcal{L}=\mathcal{L}_0+\lambda \mathcal{L}_1+\lambda^2\mathcal{L}_2+\cdots.
\end{equation}
Suppose that among the terms which appear in the $\mathcal{O}(\lambda^n)$ contribution, there is a term, $\mathcal{L}_n^{R}$, which vanishes when the fields satisfy the equations of motion for the lowest-order Lagrangian $\mathcal{L}_0$.  In this case, we can by integration by parts always write
\begin{equation} \mathcal{L}_n^{R}\simeq f^i\left([\phi]\right)\frac{\delta^{\rm EL}\mathcal{L}_0}{\delta \phi^i}.\end{equation}
Here $\simeq$ means equality up to a total derivative and $f^i$ is some function with $[\phi]$ standing for dependence on the fields, their derivatives and possibly the coordinates.

Any such interaction $\mathcal{L}_n^{R}$ can be removed, without altering any of the other terms at lower or equal order, by performing a field redefinition.  The required redefinition is
\begin{equation}\phi^i\rightarrow  \phi^i-\lambda^n f^i\left([ \phi]\right).\end{equation}
Under this redefinition, the  $\mathcal{O}\left(\lambda^n \right)$ terms in the action change as
\begin{equation} \lambda^n \mathcal{L}_n\rightarrow \lambda^n \mathcal{L}_n- \lambda^n f^i\left([\phi]\right)\frac{\delta^{\rm EL}\mathcal{L}_0}{\delta \phi^i}+\mathcal{O}\left(\lambda^{n+1}\right).\end{equation}
Hence, to  $\mathcal{O}\left(\lambda^n \right)$, the only effect of this redefinition is to cancel $\mathcal{L}_n^{R}$.

\subsection*{The DBI Action:}

We will, for simplicity, present our analysis within the context of a flat bulk space. However, it is straightforward to prove that all arguments go through with minimal modifications even for a curved bulk---such as AdS spacetime---and that the final result does not change.
In the gauge \eqref{146}, the most general action for a $d=4$ brane embedded in a $d=5$ bulk space is
\begin{equation}
\label{generalaction} 
S=\left.
\int d^4x\ \sqrt{-g}{\mathcal{L}}\left(g_{\mu\nu},\nabla_\mu,R^{\rho}_{\ \sigma\mu\nu},K_{\mu\nu}\right)
\right|_{g_{\mu\nu}=\eta_{\mu\nu}+\partial_\mu \pi\partial_\nu\pi} \ .
\end{equation}
where $\mu,\nu,$ $\dots$$ =0,\dots,3$.
This will be a power series in some length scale $l$,  which plays the role of $\lambda$ above.  $K$ and $\nabla$ get one power of $l$, and $R$ gets two powers.  
 We can use the Gauss-Codazzi relation,
\begin{equation} R_{\mu\nu\rho\sigma}-K_{\mu \rho}K_{\nu \sigma}+K_{\mu \sigma}K_{\nu \rho}=0,\end{equation}
to eliminate all occurrences of $R$ in favor of $K$, so the action becomes
\begin{equation}\label{generalactionK}
S=\left. \int d^4x\ \sqrt{-g}{\mathcal{L}}\left(g_{\mu\nu},\nabla_\mu,K_{\mu\nu}\right)\right|_{g_{\mu\nu}=\eta_{\mu\nu}+\partial_\mu \pi\partial_\nu\pi} \ .
\end{equation}

Let us analyze the case where there are no $\nabla$ operators.  Then the most general Lagrangian can be written in the form
\begin{equation}{\mathcal{L}}=M^4\left( \mathcal{L}_0+l  \mathcal{L}_1+l^2  \mathcal{L}_2+l^3  \mathcal{L}_3+\cdots \right) \ , \label{new1}\end{equation}
where
\begin{eqnarray}
\mathcal{L}_0&=&A_1, \nonumber \\
\mathcal{L}_1&=&A_2 \left[ K \right], \nonumber \\
\mathcal{L}_2&=&B_1 \left[ K^2\right]+B_2\left[ K\right]^2,  \nonumber\\ 
\mathcal{L}_3&=&C_1\left[ K^3\right]+C_2 \left[ K^2\right] \left[ K\right] +C_3\left[ K \right] ^3, \nonumber\\ 
\mathcal{L}_4&=& D_1\left[ K^4\right]+D_2 \left[ K^3\right] \left[ K\right] +D_3\left[ K^2 \right]^2 +D_4\left[ K^2\right]\left[ K\right]^2+D_5\left[ K\right]^4,\nonumber \\ 
\mathcal{L}_5&=& F_1\left[ K^5\right]+F_2 \left[ K^4\right] \left[ K\right] +F_3\left[ K^3 \right] \left[ K\right]^2 +F_4\left[ K^3\right]\left[ K^2\right]  \nonumber \\ 
&&+F_5\left[ K^2\right]^2\left[ K\right]+F_6\left[ K^2\right]\left[ K\right]^3+F_7\left[ K\right]^5, \nonumber\\
&\vdots&  \label{U5equation}
\end{eqnarray}
These are simply all possible contractions of $K_{\mu\nu}$.  The square bracket indicates a trace with indices raised with $g^{\mu\nu}$ ---that is,  $[K] = g^{\mu\nu} K_{\mu\nu}$, $[K^2] = g^{\mu\alpha} K_{\alpha\beta}g^{\beta\nu}K_{\nu\mu}$, and so on.  The coefficients are generic dimensionless parameters.  
 
 \subsection*{The $\pi$ Field Redefinition:}

The zero-th order Lagrangian is
\begin{equation}\mathcal{L}_0 = \sqrt{-g}= \sqrt{1+(\partial\pi)^2}.\end{equation}
This leads to the zero-th order equation of motion 
\begin{equation}\frac{\delta^{\rm EL}\mathcal{L}_0}{\delta \pi}=-\gamma\square \pi+\gamma^3 \partial^\mu\pi\partial^\nu\pi\partial_\mu\partial_\nu\pi=[K],\end{equation}
with $\gamma\equiv{1/ \sqrt{1+(\partial\pi)^2}}.$
Therefore, the lowest order equation of motion is simply $[K]$ itself, so any term proportional to trace of $K_{\mu\nu}$ can be eliminated by field redefinition.  
The only terms in a general ${\mathcal{L}}_{n}$ that are not of this form are the cyclic traces $\left[ K^n\right]$, of which there is only one at each order $n$.  All the other terms are proportional to $[K]$ and their coefficients are adjustable. 
 
Now, at every order, there is a special contraction of $K$'s,
\begin{eqnarray} \mathcal{L}_1^{\rm G}(K) &=&[K], \nonumber\\ 
 \mathcal{L}_2^{\rm G}(K) &=&[K]^2-[K ^2], \nonumber \\
\mathcal{L}_3^{\rm G} (K)&=& [K]^3-3 [K][K ^2]+2[K ^3] , \nonumber\\
\mathcal{L}_4^{\rm G} (K)&=&[K]^4
-6[K ^2][K]^2+8[K ^3][K]+3[K ^2]^2 -6[K ^4] ,\nonumber \\
&\vdots& \label{combos}
\end{eqnarray}
called the ``characteristic polynomials''. These are terms in the expansion of the determinant of $1+K$ in powers of $K$,
\begin{equation}\det(1+ K)=1+\mathcal{L}_1^{\rm G}(K)+\frac{1}{2} \mathcal{L}_2^{\rm G}(K) +\frac{1}{3!} \mathcal{L}_3^{\rm G}(K)+\frac{1}{4!} \mathcal{L}_4^{\rm G}(K)+\cdots \end{equation}
The terms $ \mathcal{L}_n^{\rm G}(K)$ are precisely the Galileons for $n<4$, are a total derivative when $n=4$ and vanish identically when $n>4$. 
They can be written explicitly as \cite{Trodden:2011xh}
\begin{equation} \mathcal{L}_n^{\rm G}(K)=\sum_p\left(-1\right)^{p}\eta^{\mu_1p(\nu_1)}\eta^{\mu_2p(\nu_2)}\cdots\eta^{\mu_np(\nu_n)} \left(K_{\mu_1\nu_1}K_{\mu_2\nu_2}\cdots K_{\mu_n\nu_n}\right).\end{equation}
The sum is over all permutations of the $\nu$ indices, with $(-1)^p$ the sign of the permutation.  

We see that the coefficient of $\left[ K^n\right]$ in these special combinations is non-vanishing at each order.  Thus, by using a field redefinition of $\pi$ to adjust the coefficients of the terms (\ref{U5equation}) which are proportional to $[K]$, one can bring each of the Lagrangians $\mathcal{L}_n$ into the form of the combinations (\ref{combos}).  After this $\pi$ redefinition, the action has a finite number of terms---precisely the four Galileons.  Therefore, to all orders in $l$, we have
\begin{equation}\label{galaction} {\mathcal{L}}=M^4\left( a_0 \mathcal{L}_0^{\rm G}+a_1 l \mathcal{L}_1^{\rm G}+a_2 l^2 \mathcal{L}_2^{\rm G}+a_{3}l^{3}\mathcal{L}_{3}^{\rm G} \right). \end{equation}
The coefficients $a_0\cdots,a_3$ are now complicated functions of the original coefficients.  Note that the coefficient $a_1$ remains adjustable.
We remind the reader that we have proven \eqref{galaction} for the case when $\mathcal{L}$ in \eqref{generalactionK} is independent of $\nabla$. 

Now return to the possibility of having derivatives $\nabla$ in the action (\ref{generalactionK}).  Note that derivatives always have to come in pairs.  Up  through order $l^3$, there is no way to write any terms involving derivatives which is not a total derivative.  Thus (\ref{galaction}) is accurate up through order $l^3$, even when including the possibility of derivatives.  

At order $l^4$ and above, there seems to be a problem because there can be terms such as $K_{\mu\nu}\square K^{\mu\nu}$. These are not proportional to $[K]$ and, hence, can not be eliminated by the zero-th order equations.  It could be, however, there are enough combinations like (\ref{combos}), but now involving derivatives, into which these terms could be placed so that they become total derivatives.
We leave a study of these higher-order $\nabla$ terms for future work.


\section*{Acknowledgments}
We wish to thank Ray Volkas for discussions, and Kurt Hinterbichler for discussions and early collaboration, especially with the arguments of the appendix. The work of Burt Ovrut is supported in part by the DOE under contract No. DE-AC02-76-ER-03071.  B.A.O. acknowledges partial support from the NSF RTG grant DMS-0636606 and from NSF Grant 555913/14 for International Collaboration. J.S. is supported by NASA ATP grant NNX08AH27G, and funds provided by the University of Pennsylvania.


\end{document}